\documentclass[twocolumn,floatfix,prb,aps,showpacs,bibliography]{revtex4-1}
\usepackage{graphicx,amsmath,amssymb,color}
\usepackage{nicefrac}
\usepackage{url}
\usepackage[titletoc,title]{appendix}
\usepackage{color,amsmath,amssymb,amsfonts,graphicx,tabularx}
\usepackage{nicefrac}
\usepackage[titletoc,title]{appendix}
\usepackage[unicode=true,colorlinks=true]{hyperref}
\hypersetup{linkcolor=blue,citecolor=blue,urlcolor=blue}
\makeatletter
\newcommand*{\rom}[1]{\expandafter\@slowromancap\romannumeral #1@}
\makeatother

\newcommand{\be}{\begin{equation}}
\newcommand{\ee}{\end{equation}}

\newcommand{\ba}{\begin{eqnarray}}
\newcommand{\ea}{\end{eqnarray}}

\renewcommand{\vec}[1]{{\textbf{\textit{#1}}}}

\begin{document}

\title{Fermi wave vector for the non-fully spin polarized composite-fermion Fermi sea}
\author{Ajit C. Balram$^{1}$ and J. K. Jain$^{2}$}
\affiliation{
   $^{1}$Niels Bohr International Academy and the Center for Quantum Devices, Niels Bohr Institute, University of Copenhagen, 2100 Copenhagen, Denmark}
\affiliation{
   $^{2}$Department of Physics, 104 Davey Lab, Pennsylvania State University, University Park, Pennsylvania 16802, USA}
\date{\today}

\begin{abstract} The fully spin polarized composite fermion (CF) Fermi sea at half filled lowest Landau level has a Fermi wave vector $k^*_{\rm F}=\sqrt{4\pi\rho_e}$, where $\rho_e$ is the density of electrons or composite fermions, supporting the notion that the interaction between composite fermions can be treated perturbatively. Away from $\nu=1/2$, the area is seen to be consistent with $k^*_{\rm F}=\sqrt{4\pi\rho_e}$ for $\nu<1/2$ but $k^*_{\rm F}=\sqrt{4\pi\rho_h}$ for $\nu>1/2$, where $\rho_h$ is the density of holes in the lowest Landau level. This result is consistent with particle-hole symmetry in the lowest Landau level. We investigate in this article the Fermi wave vector of the spin-singlet CF Fermi sea (CFFS) at $\nu=1/2$, for which particle-hole symmetry is not a consideration. Using the microscopic CF theory, we find that for the spin-singlet CFFS the Fermi wave vectors for up and down spin CFFSs at $\nu=1/2$ are consistent with $k^{*\uparrow,\downarrow}_{\rm F}=\sqrt{4\pi\rho^{\uparrow,\downarrow}_e}$, where $\rho^{\uparrow}_e=\rho^{\downarrow}_e=\rho_e/2$,
%=k^{*\downarrow}_{\rm F}$, 
which implies that the residual interactions between composite fermions do not cause a non-perturbative correction for non-fully spin polarized CFFS either. Our results suggest the natural conjecture that for arbitrary spin polarization the CF Fermi wave vectors are given by $k^{*\uparrow}_{\rm F}=\sqrt{4\pi\rho^{\uparrow}_e}$ and $k^{*\downarrow}_{\rm F}=\sqrt{4\pi\rho^{\downarrow}_e}$. 
\end{abstract}
\maketitle

\section{Introduction}

The emergence of a Fermi sea at half filled Landau level (LL)\cite{Halperin93,Kalmeyer92,Willett93,Kang93,Goldman94,Smet96,Smet98,Willett99,Smet99,Freytag02} is remarkable given that the original Hamiltonian has no kinetic energy. Its appearance is a consequence of the formation of composite fermions\cite{Jain89,Jain07}, which experience, on average, no magnetic field at a filling factor $\nu=1/2$. In recent years, accurate determination of the Fermi wave vector of composite fermions  from commensurability oscillations at and near half filled Landau level (LL)\cite{Kamburov12,Kamburov14,Kamburov14b,Kamburov14c,Mueed15} has shed important new light into the physics of the composite fermion (CF) Fermi sea. At $\nu=1/2$, the measured Fermi wave vector of composite fermions, denoted by $k_{\rm F}^{*}$, is found to be $k_{\rm F}^{*}=\sqrt{4\pi\rho_e}$, consistent with a fully spin polarized Fermi sea of composite fermions with density $\rho_e$. Slightly away from $\nu=1/2$, the CF Fermi wave vector has been found to be consistent with that of the minority charge carriers in the lowest LL (LLL)\cite{Kamburov14}, i.e., $k_{\rm F}^{*}=\sqrt{4\pi\rho_e}$ for $\nu<1/2$ and $k_{\rm F}^{*}=\sqrt{4\pi\rho_h}$ for $\nu>1/2$, where $\rho_h$ is the density of holes in the LLL. The dimensionless quantity $k_{\rm F}^{*}\ell$, where $\ell=\sqrt{\hbar c/eB}$ is the magnetic length, is given by $k_{\rm F}^{*}\ell=\sqrt{2\nu}$ for $\nu<1/2$ and $k_{\rm F}^{*}\ell=\sqrt{2(1-\nu)}$ for $\nu>1/2$.

This experimental observation has put focus on the following paradox. At first glance, the CF theory seems to offer two distinct choices for any given filling factor. The state can be described in terms of composite fermions formed from binding of vortices to electrons, which have the same density as electrons, namely $\rho_e$. Alternatively, the state can be described in terms of composite fermions formed from binding of vortices to holes in the LLL, which have the density $\rho_h$. Are these two different states of matter? This point of view has been taken in Ref.~\cite{Barkeshli15}, which investigates the consequences of a spontaneous breaking of the particle-hole (PH) symmetry in the LLL. However, numerical calculations strongly suggest that, in spite of the seemingly different physics, the states formed from composite fermions made of electrons and composite fermions made of holes are ultimately dual descriptions of the same state. For example, the CFFS at $\nu=1/2$ constructed from composite-fermionizing electrons has been found to be essentially identical to that constructed from composite-fermionizing holes\cite{Rezayi00}. As another example, the states at $\nu=(n+1)/(2n+1)$ can be constructed either as $\nu^*=n$ of composite fermions formed from holes or $\nu^*=n+1$ of composite fermions made of electrons (in a negative magnetic field); both of these descriptions produce identical quantum numbers for the ground and low energy excited states, and, indeed, their actual wave functions have  close to 100\% overlap for finite but not very small systems\cite{Wu93,Balram16b}. The PH symmetry of composite fermions has motivated a Dirac CF theory that builds PH symmetry in a manifest fashion\cite{Son15}. The Chern-Simons field theory approach\cite{Lopez91,Halperin93} is not projected into the LLL, and thus does not allow a consideration of composite fermions formed from binding of vortices to the holes of the LLL, but has nonetheless been shown\cite{Wang17} to produce results consistent with PH symmetry to certain nontrivial orders in perturbation theory. For experimental parameters, the positions of the commensurability oscillations minima predicted from the Dirac CF theory and the Chern-Simons field theory agree to a high degree\cite{Cheung17}.

Even if one assumes PH symmetry, the question ``Is the $k^*_{\rm F}$ determined by $\rho_e$, by $\rho_h$, or by something else?" still remains. At $\nu=1/2$, where $\rho_e=\rho_h$, one would expect $k_{\rm F}^{*}=\sqrt{4\pi\rho_e}$, if one assumes that the residual interactions between composite fermions can be treated perturbatively. However, away from $\nu=1/2$ the answer is not obvious. In Ref.~\onlinecite{Balram15b}, we addressed this question in an unbiased fashion starting from the microscopic wave functions of composite fermions, which are known to be very close to the exact solutions of the Coulomb problem\cite{Dev92,Jain97,Jain07,Balram13,Jain14,Balram15a}. Following an earlier work\cite{Kamilla97}, we used the Friedel oscillations in the pair correlation function $g(r)$, which is equal to the normalized probability of finding two particles a distance $r$ apart from each other in the ground state, to determine the CF Fermi wave vector at $\nu=1/2$ and nearby filling factors, and found it to be close to that given by the minority carrier rule. (In the Appendix~\ref{app:fully_polarized} we show results for fully spin polarized systems for higher particle numbers than those given in Ref.~\onlinecite{Balram15b}, which bring $k^*_{\rm F}$ into better agreement with the value ascertained from the minority carrier rule. We also obtain the static structure factor\cite{Giuliani08,Kamilla97} for fully spin polarized systems and compare it with the predictions from topological field theories\cite{Gromov15,Nguyen17} and the Dirac composite fermion approach\cite{Son15,Nguyen17,Nguyen17b}.) For this purpose, we fitted the oscillations in $g(r)$ at an intermediate range of $r$, because for small $r$ the short distance correlations dominate, whereas for large $r$ the pair correlation function exponentially approaches unity due to the gap.  We stress that no Fermi wave vector was built into the initial wave function. It is notable that well defined oscillations occur at intermediate distances at $\nu=n/(2n\pm 1)$ even for moderate values of $n$\cite{Kamilla97,Balram15b}. We also showed that the result was independent of whether composite fermions made of electrons or holes are used.

The goal of this work is to extend the work of Ref.~\onlinecite{Balram15b} to include the spin degree of freedom and study a non-fully spin polarized CFFS at $\nu=1/2$. The reason is that PH symmetry is not relevant to a CFFS that is not fully spin polarized, because, once the spin degree of freedom is active, particle-hole symmetry relates states at $\nu$ and $2-\nu$. The apparent dichotomy that exists in the CF theory for fully polarized states around $\nu=1/2$ does not arise for systems that are not fully polarized; now only composite fermions formed from {\em electrons} see a zero effective magnetic field at filling factor $1/2$. From this perspective, one might argue that provided a Fermi sea is formed, its area must be determined by the density of up and down spin electrons, and not up and down spin holes. Nonetheless, it is in principle possible that the Luttinger theorem is violated for this state. Furthermore, as we move away from $\nu=1/2$, there is no longer a symmetry that relates the $k^*_{\rm F}\ell$'s on the two sides. 

Additionally, the question has experimental relevance. It is well known, both theoretically\cite{Park98,Toke07b,Zhang16,Zhang17} and experimentally\cite{Kukushkin99,Melinte00,Giudici08,Finck10,Eisenstein16}, that the CFFS is partially spin polarized at relatively small but experimentally attainable Zeeman energies. This is expected from the fact that the fractional quantum Hall (FQH) states at $\nu=n/(2n\pm 1)$ have, in general, several spin polarizations \cite{Eisenstein89, Eisenstein90,Engel92,Du95,Kang97,Kukushkin99,Yeh99,Kukushkin00,Freytag01,Freytag02,Tracy07,Tiemann12,Feldman13,Liu14}, which have been qualitatively and quantitatively explained by the CF theory in terms of partially spin polarized states of composite fermions\cite{Wu93,Park98,Balram15a,Zhang16}. Even a spin-singlet CFFS can be obtained by reducing the Land\'e g factor to zero by application of pressure\cite{Leadley97,Kang97} or by going to a two-valley system with zero valley splitting\cite{Bishop07,Padmanabhan09,Padmanabhan10}.

These questions have motivated us to consider the Fermi wave vector for a non-fully spin polarized CFFS.  For technical reasons, it is convenient for us to study a spin-singlet CFFS, but our results are straightforwardly generalizable to CFFS with arbitrary spin polarization. Since the technical details of this work are very similar to those of Ref.~\onlinecite{Balram15b}, it would suffice to give an outline. We evaluate the CF Fermi wave vector for spin-singlet states from Friedel oscillations within the CF framework and find that it is consistent with the density of electrons. For this purpose, we consider both the spin-singlet CFFS at $\nu=1/2$ and the spin-singlet FQH states at $\nu=n/(2n\pm 1)$ for large $n$ (only even values of $n$ produce spin-singlet states) where the densities of spin up and spin down composite fermions are given by $\rho^{\uparrow}_e=\rho^{\downarrow}_e=\rho_e/2$. We use the projected Jain wave functions for the calculation, which are known to describe the physics of the LLL very accurately\cite{Dev92,Jain97,Jain07,Balram13,Jain14,Balram15a}. At $\nu=1/2$, our calculated CF Fermi wave vector for both up and down spin composite fermions is consistent with $k^*_{\rm F}=\sqrt{2\pi \rho_e}$. Given that the model of noninteracting composite fermions is valid for both fully spin polarized and spin singlet states, we expect the Fermi wave vectors for partially spin polarized CFFS to be given by $k^{*\uparrow}_{\rm F}=\sqrt{4\pi \rho^{\uparrow}_e}$ and $k^{*\downarrow}_{\rm F}=\sqrt{4\pi \rho^{\downarrow}_e}$, where $\rho^{\uparrow}_e$ and $\rho^{\downarrow}_e$ are densities of up- and down-spin electrons or composite fermions. The situation is less clear away from $\nu=1/2$, but our calculations suggest a `tent-like' structure for $k^{*}_{\rm F}$, approaching $k^{*}_{\rm F}=\sqrt{2\pi \rho_e}$ sufficiently close to $\nu=1/2$.

\section{Background}

As a background, the FQH effect of electrons at filling factors along the Jain sequence $\nu=n/(2pn\pm1)$ is described as the integer quantum Hall (IQH) state of composite fermions carrying $2p$ vortices (denoted as $^{2p}$CFs) with $n$ filled $\Lambda$ levels ($\Lambda$Ls)\cite{Jain89}. (The term $\Lambda$Ls refers to emergent Landau-like levels of composite fermions, which reside entirely within the LLL of electrons.) We shall specialize to $2p=2$ below. For spinful electrons, we write\cite{Wu93,Park98,Jain07} the CF filling as $n=n_{\uparrow}+n_{\downarrow}$, where $n_{\uparrow}$ and $n_{\downarrow}$ are the number of filled spin-up and spin-down $\Lambda$Ls respectively. The Jain wave function\cite{Jain07} for this state is given by: 
\be
\Psi_{\frac{n}{2n\pm1}}={\cal P}_{\rm LLL} \Phi_{\pm n} J^{2}={\cal P}_{\rm LLL} \Phi_{\pm n_{\uparrow}}\Phi_{\pm n_{\downarrow}} J^{2}
\label{eq_proj_CF}
\ee
where  $J= \prod_{1\leq j<k \leq N}(z_j-z_k)$
is the Jastrow factor, $z_i$ is the $i^{\text{th}}$ electron coordinate written as a complex number in the two-dimensional plane, $\Phi_{n_\uparrow}$ ($\Phi_{n_\downarrow}$) is the Slater determinant wave function for $n_{\uparrow}$ ($n_{\downarrow}$) filled LLs of electrons, $\Phi_{-n}=[\Phi_{n}]^{*}$, and ${\cal P}_{\rm LLL}$ denotes LLL projection. In this paper we shall consider only spin-singlet states i.e., $n_{\uparrow}$=$n_{\downarrow}=n/2$. For all our calculations we shall use the Jain-Kamilla projection method, details of which can be found in the literature\cite{Jain97,Jain97b,Jain07,Moller05,Davenport12,Balram15a,Mukherjee15b}. For $n\rightarrow\infty$ the sequence $n/(2n\pm 1)$ approaches the filling factor $\nu=1/2$, where a spin-singlet CFFS is predicted to exist\cite{Toke07b,Balram15c} at zero Zeeman energy. Throughout this work we use the spherical geometry\cite{Haldane83,Greiter11} in which $N$ electrons reside on the surface of a sphere and see a radial magnetic flux of $2Qhc/e$ emanated from a Dirac monopole sitting at the center of the sphere. All states considered in this work have a uniform density on the sphere i.e., have total orbital angular momentum $L=0$, and have a total spin $S=0$.

Note that unlike for the fully polarized states around $\nu=1/2$, the CF theory gives a unique description for the states at $\nu=n/(2n\pm 1)$ involving spins.  These states are understood as IQH states of composite fermions formed from attaching vortices to {\em electrons}; these composite fermions see a positive effective magnetic field at $\nu=n/(2n+1)$ and negative effective magnetic field at $\nu=n/(2n-1)$.

\section{Calculations and results}

We extract the Fermi wave vector for composite fermions from the Fermi-sea like Friedel oscillations seen in the pair-correlation function of FQH states\cite{Rezayi94,Park98b,Kamilla97,Balram15b}. The pair-correlation function for a homogeneous non-interacting Fermi gas in zero magnetic field with equal number of up and down spins in two dimensions is given by\cite{Giuliani08}:
\begin{equation}
g_{\sigma,\sigma'}(r)=1-\delta_{\sigma,\sigma'} \Bigg( \frac{2J_{1}(rk_{\rm F})}{rk_{\rm F}}  \Bigg)^{2}
\end{equation}
where the Fermi wave vector $k_{\rm F}=\sqrt{2\pi\rho}$, $\rho$ is the total fermion density and $J_{1}(x)$ is the Bessel function of order one of the first kind. Clearly, we have $g_{\uparrow,\downarrow}(r)=1=g_{\downarrow,\uparrow}(r)$ for all $r$ as there are no Pauli correlations between non-interacting fermions of opposite spins. The oscillatory part of $g_{\sigma,\sigma}(r)$ for large $rk_{\rm F}$ goes as $(rk_{\rm F})^3\sin(2k_{\rm F}r)$, which lead to the well-known $2k_{\rm F}$ Friedel oscillations. This motivates us to define the Fermi wave vector for FQH states through the the pair correlation function, for which we assume the form\cite{Kamilla97,Balram15b}:
\be
g_{\sigma\sigma}(r)=1+A (r\sqrt{2\pi \rho_{\rm e}})^{-\alpha} \sin(2k^*_{\rm F} r+\theta)
\label{pair_correlation_fitting_form}
\ee
where $\rho_{\rm e}$ is the electron density and $A$, $\alpha$, $k^*_{\rm F}$ and $\theta$ are fitting parameters.

\begin{figure}[t]
\begin{center}
\includegraphics[width=7.5cm,height=4.5cm]{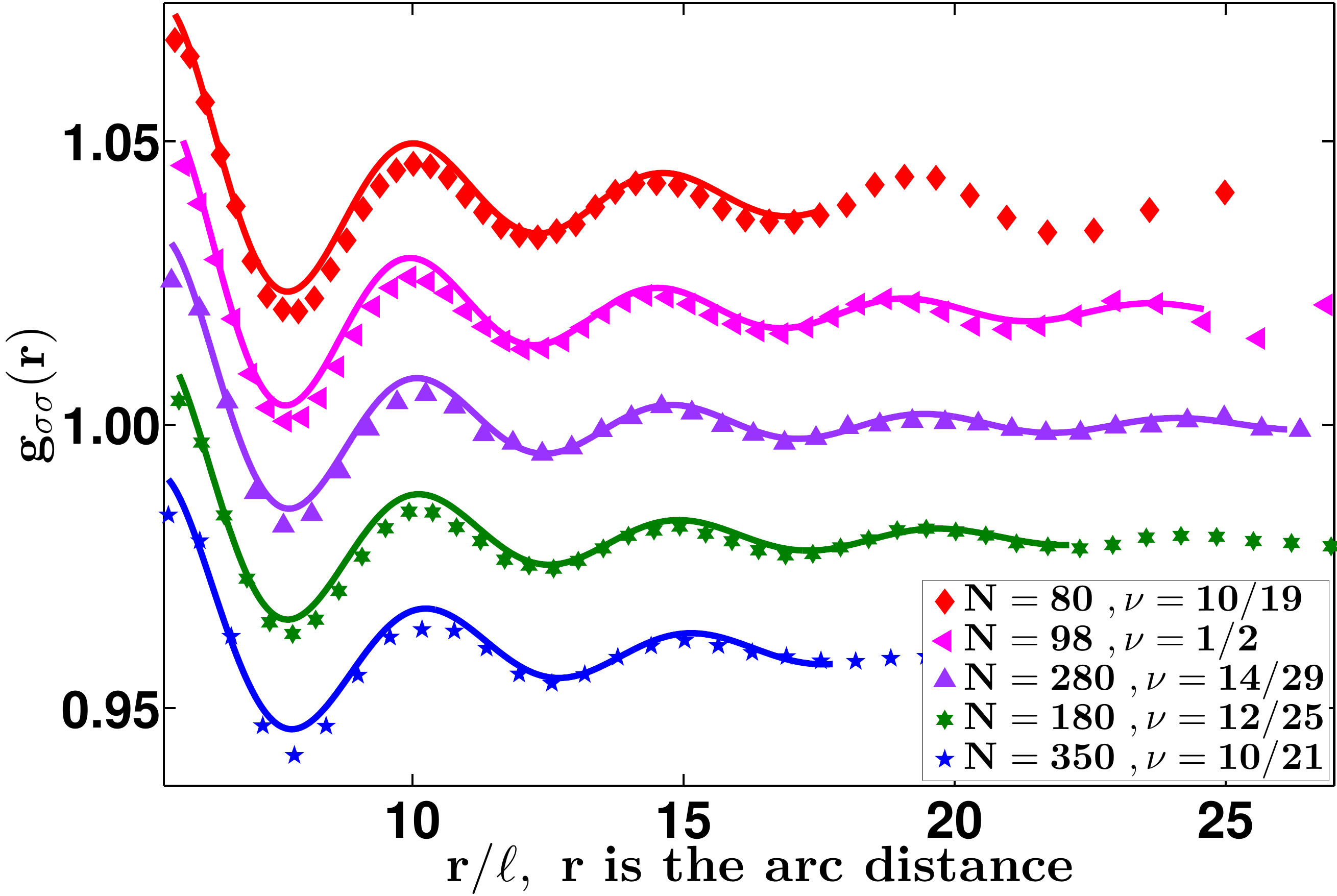}
\includegraphics[width=7.5cm,height=4.5cm]{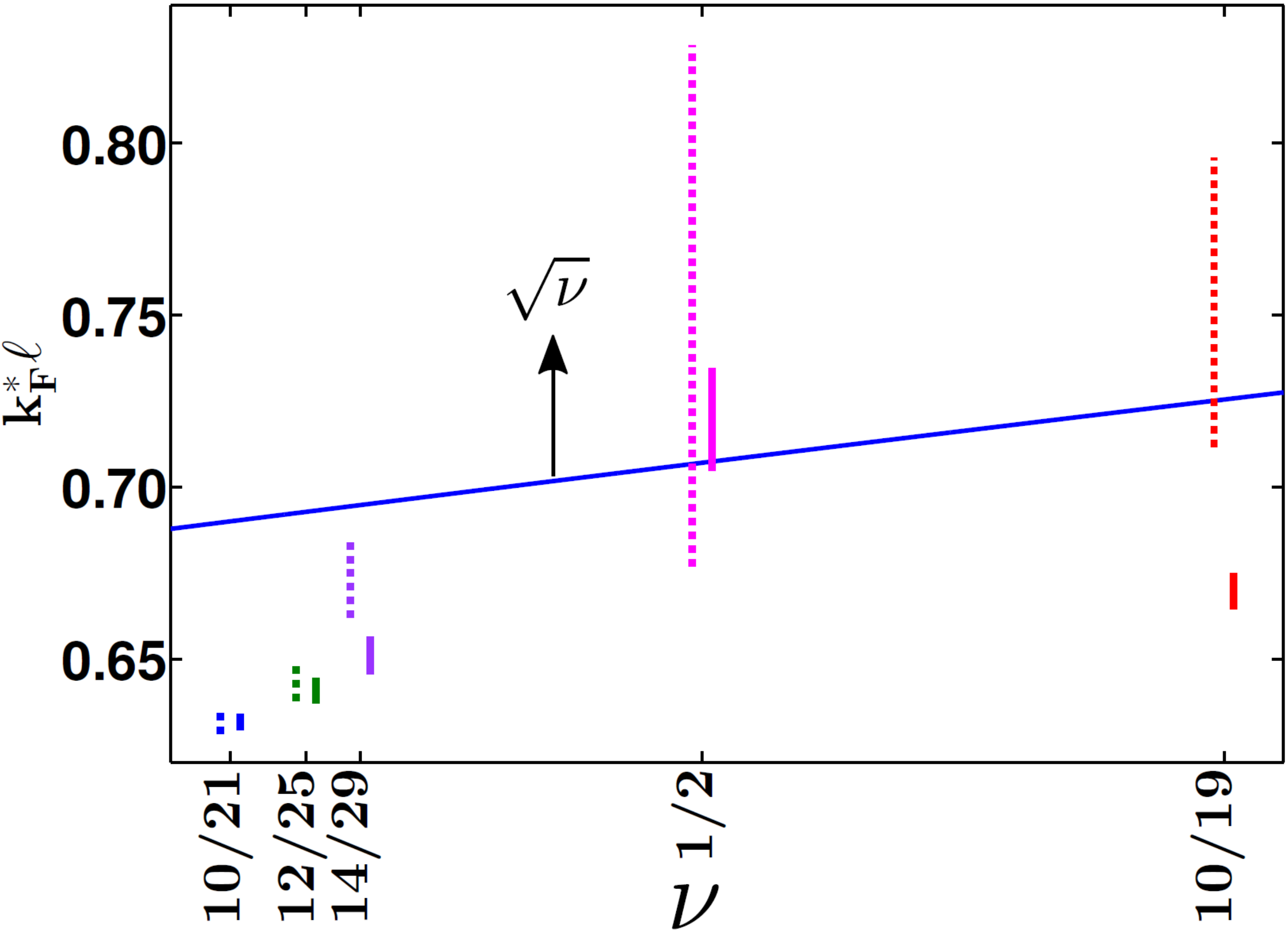}
\end{center}
\caption{(a) Pair correlation function $g(r)$ as a function of $r/\ell$, where $r$ is the arc distance on the sphere, obtained using the projected wave functions of Eq.~(\ref{eq_proj_CF}). The solid lines are fits using Eq.~(\ref{pair_correlation_fitting_form}) for oscillations in an intermediate range of $r$. The curves (except for 14/29) have been shifted up or down by multiples of 0.02 to avoid clutter. (b) Thermodynamic value of $k_{\rm F}^*\ell$ as a function of $\nu$ from arc fits (solid vertical bars) and chord fits (dashed vertical bars), slightly shifted horizontally for clarity. The mean-field value $k_{\rm F}^{\rm *MF} \ell=\sqrt{\nu}$ is shown for reference. }
\label{kF}
\end{figure}

We have evaluated the same-spin pair-correlation function $g_{\sigma,\sigma}(r)$ for $\nu=n/(2n+1)$ for up to $n=14$ using the Monte Carlo method in the spherical geometry choosing $r$ as the chord or the arc distance. The $g_{\sigma,\sigma}(r)$ for the largest systems considered in this work are shown in Fig.~\ref{kF}(a). In the same figure we also show the fits for the pair-correlation function obtained using Eq.~(\ref{pair_correlation_fitting_form}). In this fitting we have discarded small values of $rk_{\rm F}$, where short distance physics is important and large values of $rk_{\rm F}$ where curvature effects become significant and oscillations decay exponentially due to the gap. For states in the sequence $n/(2n+1)$, it is possible to perform calculations for very large spin-singlet systems at filling factors slightly below $\nu=1/2$. We have carried out these calculations for up to $N=350$ electrons and have extrapolated $k_{\rm F}^{*}\ell$ to the thermodynamic limit from our finite system results. The largest system available at $\nu=1/2$ is relatively small with 98 particles, hence leading to a larger uncertainty in the evaluation of $k^*_{\rm F}$ at $\nu=1/2$ (see below). These extrapolations are shown in the Appendix~\ref{app:extrap_kF}. The reader may note that we are able to go to much larger value of $n$ and $N$ than for fully spin polarized composite fermions. The limitation in this respect is set by the $\Lambda$L index; it becomes increasingly more computationally time consuming to fill higher and higher $\Lambda$Ls. Since we need to fill only $n/2$ $\Lambda$Ls for spin singlet states, we can access high values of $n$.

For $\nu=1/2$ we calculate the Fermi wave vector by considering states of filled-shell CF systems at zero effective magnetic field\cite{Rezayi94}. This state can also be viewed as the $n\rightarrow\infty$ limit of the $n/(2n\pm 1)$ sequence.

We obtain thermodynamic extrapolation for the value of $k^*_{\rm F}\ell$ by fitting our pair correlation function to both arc and chord distances, and the results are shown separately in Fig.~\ref{kF}(b). Both should give the same result in the thermodynamic limit, but seen from the error bars, the fits from the arc distance are more accurate for the finite systems accessible to our calculations. 

An important result of our calculations is that the extrapolated value of the Fermi wave vector at $\nu=1/2$ is consistent with $k^*_{\rm F}=\sqrt{2\pi\rho_e}$ (i.e. $k^*_{\rm F}\ell=\sqrt{\nu}$), as expected in a model that assumes composite fermions to be noninteracting. It is stressed that we do not make any assumption regarding the interaction between composite fermions; in fact, the wave functions representing a strongly correlated liquid of electrons include all effects of interaction and are very accurate representations of the exact states. We also note, parenthetically, that we have also tried to calculate the Fermi wave vector at $\nu=1/2$ in the torus geometry (we thank Csaba T\H{o}ke in this regard). These calculations show that finite-size effects are very significant in the torus geometry, probably because we are forced to model the Fermi sea with a small section of the reciprocal lattice, which only crudely resembles a circle. A simple linear or polynomial extrapolation of the system size dependence of the wave vector of Friedel oscillations does not fit the numerical data points well; and the improvement of individual points by longer Monte Carlo runs may change the extrapolated value significantly. These considerations lead us to conclude that the torus data do not conclusively identify the thermodynamic limit of Fermi wave vector and hence we do not show results from torus geometry here.

At filling factors away from $1/2$, the calculated Fermi wave vector appears to approach the value $k^*_{\rm F}\ell=\sqrt{1/2}$ from both sides. Far from 1/2, our calculated $k^*_{\rm F}\ell$ deviates from $k^*_{\rm F}\ell=\sqrt{\nu}$. The presumably more accurate arc results suggest that $k^*_{\rm F}\ell$ has a tent-like shape, but this result is not fully corroborated by the chord extrapolations. Furthermore, as we go far from $\nu=1/2$, $g(r)$ has very few oscillations, and it is unclear how meaningful the concept of Fermi wave vector remains. 

A technical comment is in order. To observe the same number of oscillations as for fully spin polarized states, the system size needs to be doubled, which is not always feasible. For this reason, our calculation on spinful systems is more sensitive to finite size effects than the results of Ref.~\onlinecite{Balram15b}. The problem is especially severe for states with reverse flux attachment, where the LLL projection suffers from numerical precision issues\cite{Davenport12}. This requires us to store all quantities to a high precision, which considerably slows down the calculation for large $N$. Due to this technical issue, when we approach the CFFS from above, our extrapolations are based on smaller systems (and hence are less reliable) than as we approach it from below. Due to this reason we do not have enough accuracy for $\nu>1/2$. Nonetheless, the results are consistent with $k^*_{\rm F}=\sqrt{2\pi\rho_e}$. 

\section{Concluding remarks}

We end the article with several observations.

Given that the prediction from a model of non-interacting composite fermions is valid for the spin-singlet and fully spin polarized CFFSs, it is natural to expect that it remains valid for partially spin polarized CFFS as well. As noted above, a partially spin polarized CFFS will likely produce two different Fermi wave vectors, given by $k^{*\uparrow}_{\rm F}=\sqrt{4\pi \rho^{\uparrow}_e}$ and $k^{*\downarrow}_{\rm F}=\sqrt{4\pi \rho^{\downarrow}_e}$. There is preliminary experimental evidence\cite{Kamburov14c} for the reduction of the Fermi wave vector from its fully spin polarized value.

Ref.~\onlinecite{Balram15b} showed that the Fermi wave vectors of fully spin polarized states related by particle-hole symmetry, when measured in units of the inverse magnetic length, are identical. In other words,  we have $(k_{\rm F}^{*}\ell)_{1-\nu}=(k_{\rm F}^{*}\ell)_{\nu}$. By the same token, for partially spin polarized CFFS, we have $(k_{\rm F}^{*}\ell)_{2-\nu}=(k_{\rm F}^{*}\ell)_{\nu}$, and our results apply to the Fermi wave vector of partially spin polarized states in the vicinity of $\nu=3/2$. 

\begin{figure}[t]
\begin{center}
\includegraphics[width=7.5cm,height=4.5cm]{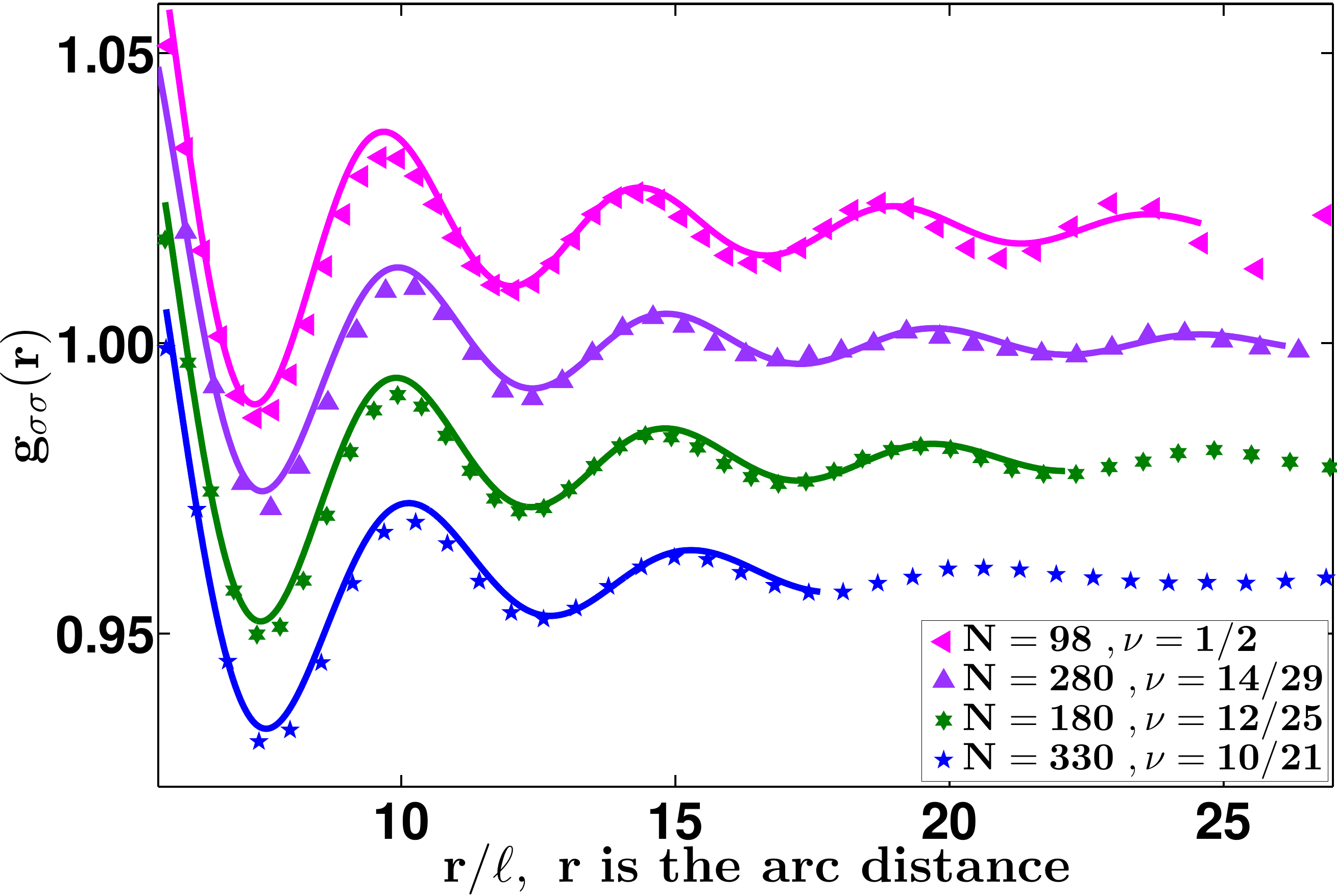}
\includegraphics[width=7.5cm,height=4.5cm]{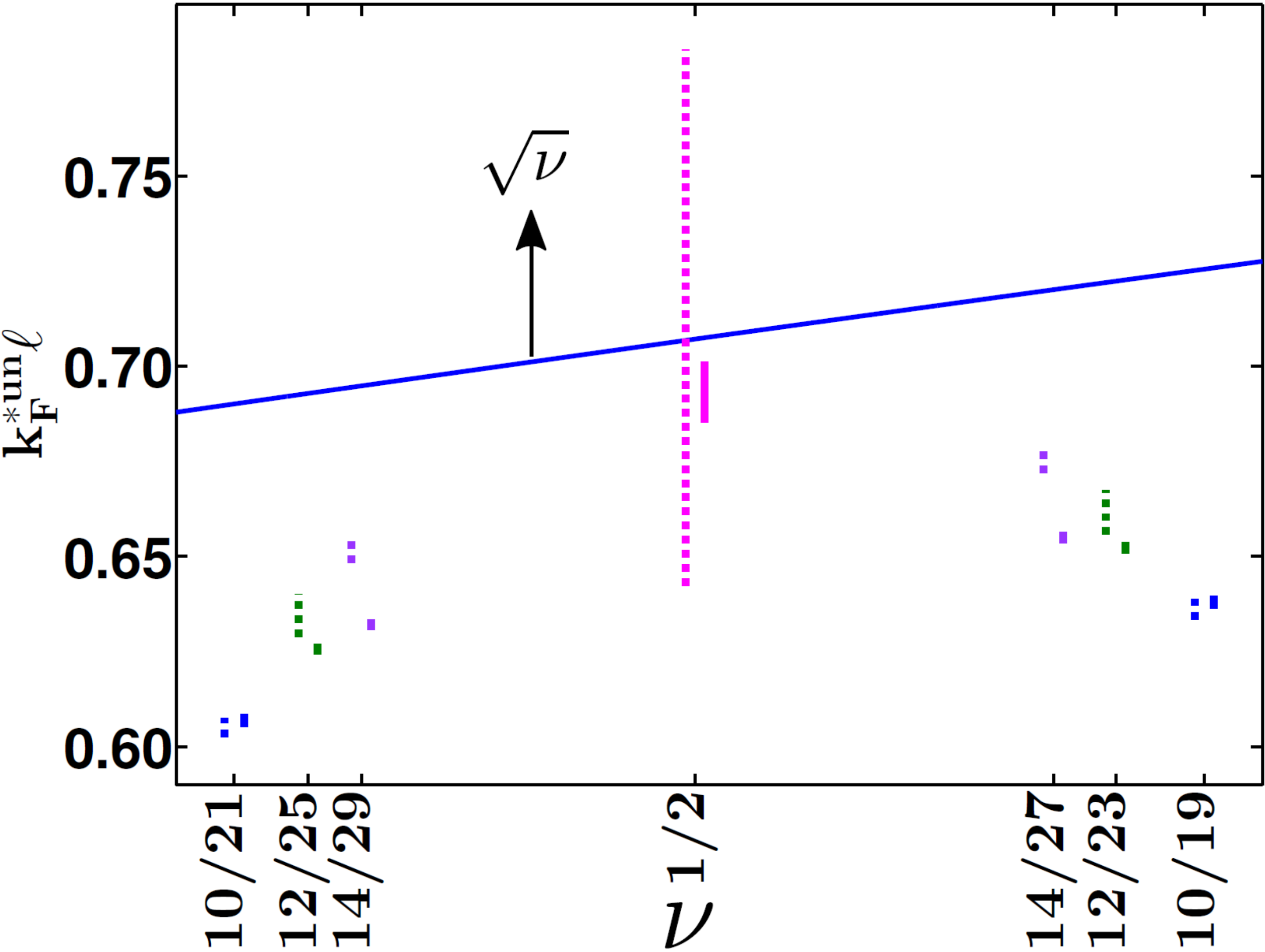}
\end{center}
\caption{Same as in Fig.~\ref{kF} but for the unprojected wave functions. Also shown for reference is the mean field value $k_{\rm F}^{\rm *MF} \ell=\sqrt{\nu}$ corresponding to composite fermions made from electrons.}
\label{kFun}
\end{figure}

We also investigate the issue of how robust the Fermi wave vector is to LL mixing. A realistic treatment of LL mixing is a complicated problem, and optimally it would require a fixed phase diffusion Monte Carlo study with the projected Jain wave functions in Eq.~(\ref{eq_proj_CF}) used as trial functions for fixing the phase\cite{Guclu05,Melik-Alaverdian95}. That is beyond the scope of this work. To gain some qualitative insight, we have calculated the Fermi wave vector from the unprojected Jain wave functions: 
\be
\Psi^{\rm un}_{\frac{n}{2n\pm1}}=\Phi_{\pm n} J^{2}=\Phi_{\pm n_{\uparrow}}\Phi_{\pm n_{\downarrow}} J^{2}
\ee
These wave functions have a small amplitude in higher LLs\cite{Trivedi91,Kamilla97b,Jain07} and are likely adiabatically connected to the projected wave functions\cite{Rezayi91}. For these wave functions, the pair correlation function at filling factors $\nu=n/(2n-1)$ and $\nu=n/(2n+1)$ are the same for a given $N$ when plotted in units of the the radius of the sphere. For a finite spin-singlet system this gives the relation
$${(k^{\rm * un}_{\rm F} \ell)_{n\over 2n-1} \over (k^{\rm * un}_{\rm F} \ell)_{n \over 2n+1}} = \sqrt{ {N-1+Q^{*} \over N-1-Q^{*}}} ,~Q^{*}= {2N-n^2 \over 4n}$$
which in the thermodynamic limit implies:
\be
(k^{\rm * un}_{\rm F} \ell)_{n\over 2n-1} = \left({2n +1 \over 2n-1}\right)^{1/2} (k^{\rm * un}_{\rm F} \ell)_{n \over 2n+1}
\label{kFrelation_un}
\ee
The estimated values of $k^{\rm * un}_{\rm F} \ell$ for the unprojected states are shown in Fig.~\ref{kFun}(b). These numbers show that to the extent it can be defined, the CFFS area away from $\nu=1/2$ depends on LL mixing. We stress that this calculation does not represent a realistic treatment of LL mixing. 
Another class of wave functions with amplitude in higher LLs is the Chern-Simons mean field wave function\cite{Jain89,Halperin93}:
\be
\Psi^{\rm CS-MF}_{\frac{n}{2n\pm1}}=\Phi_{\pm n}(B^{*})\frac{J^{2}}{|J|^{2}}
\label{eq_unproj_CF}
\ee
where $B^{*}=B-2\rho\phi_{0}$. Since the absolute value of these wave functions is the same as that of the IQH state $\Phi_{\pm n}(B^{*})$, we have $k^{\rm * MF}_{\rm F} =\sqrt{2\pi\rho_{e}}$ for all $\nu=n/(2n\pm 1)$, where $\rho_{e}$ is the density of electrons. 

For completeness, we have also calculated the Coulomb interaction energies of the spin singlet FQH states. These are tabulated in Appendix~\ref{app:gs_Coulomb_energies}.

In summary, we have shown that the Fermi wave vector for composite fermions for a spin singlet Fermi sea at $\nu=1/2$ is consistent with the prediction of the model that takes composite fermions as non-interacting. This implies that the residual interaction between composite fermion does not cause, within the accuracy of our calculations, any non-perturbative corrections to the spin-singlet CF Fermi sea. Away from $\nu=1/2$, our calculations admit the possibility of a tent-like structure for $k^*_{\rm F}\ell$.

\begin{acknowledgments}
We thank M. Mulligan, T. Senthil and D. Son for useful communications, and C. T\"oke for help with computer calculations and useful discussions. ACB was supported in part by the European Research Council (ERC) under the European Union Horizon 2020 Research and Innovation Programme, Grant Agreement No. 678862; by the Villum Foundation; and by The Center for Quantum Devices funded by the Danish National Research Foundation. JKJ was supported by the U. S. National Science Foundation Grant no. DMR-1401636. Some calculations were performed with Advanced CyberInfrastructure computational resources provided by The Institute for CyberScience at The Pennsylvania State University.
\end{acknowledgments}

\break

\begin{appendices}

\section{Extrapolation of the Fermi wave vector in the spherical geometry}
\label{app:extrap_kF}

\begin{figure*}[t]
\begin{center}
\includegraphics[width=7.5cm,height=4.5cm]{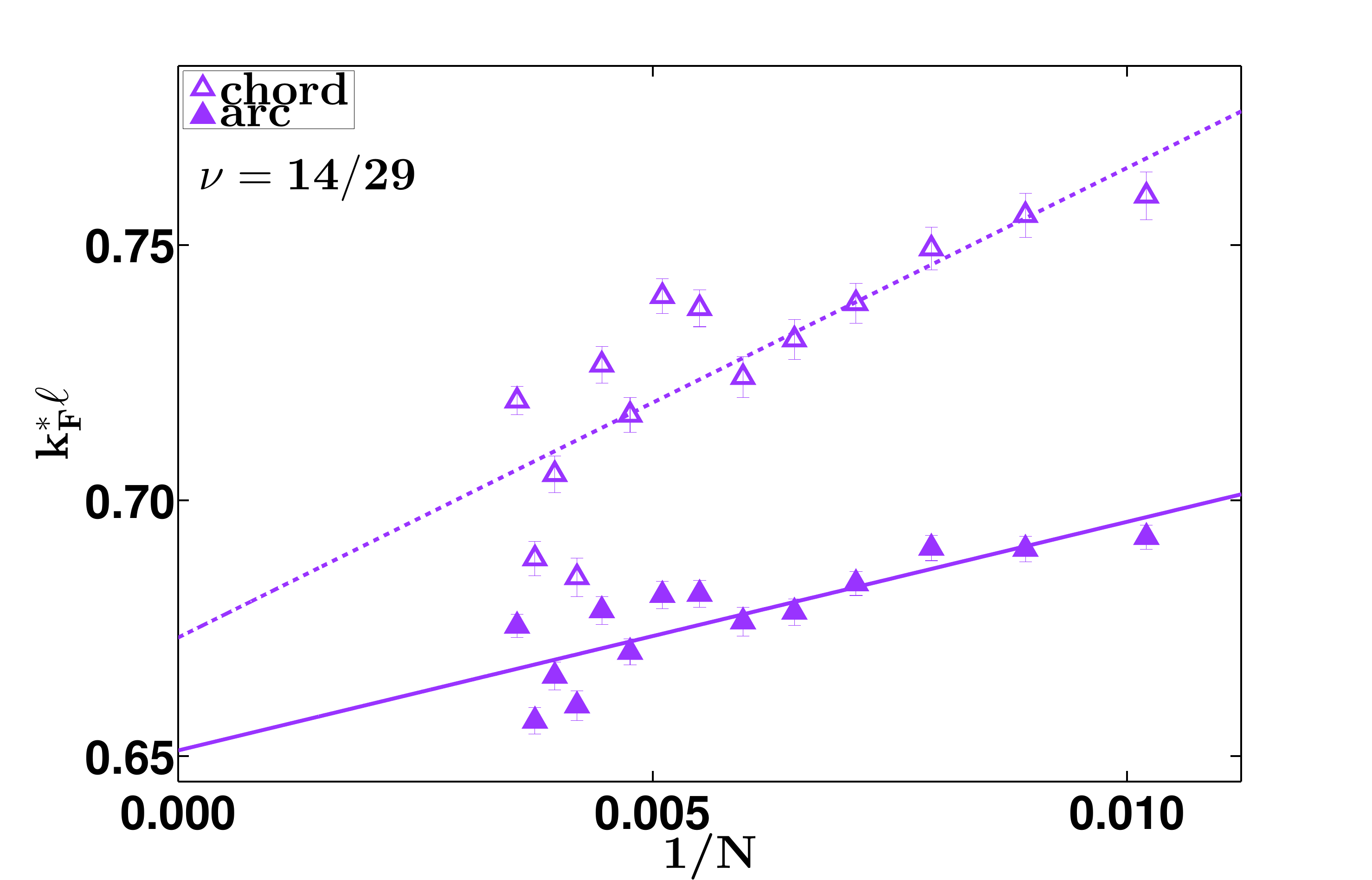} \includegraphics[width=7.5cm,height=4.5cm]{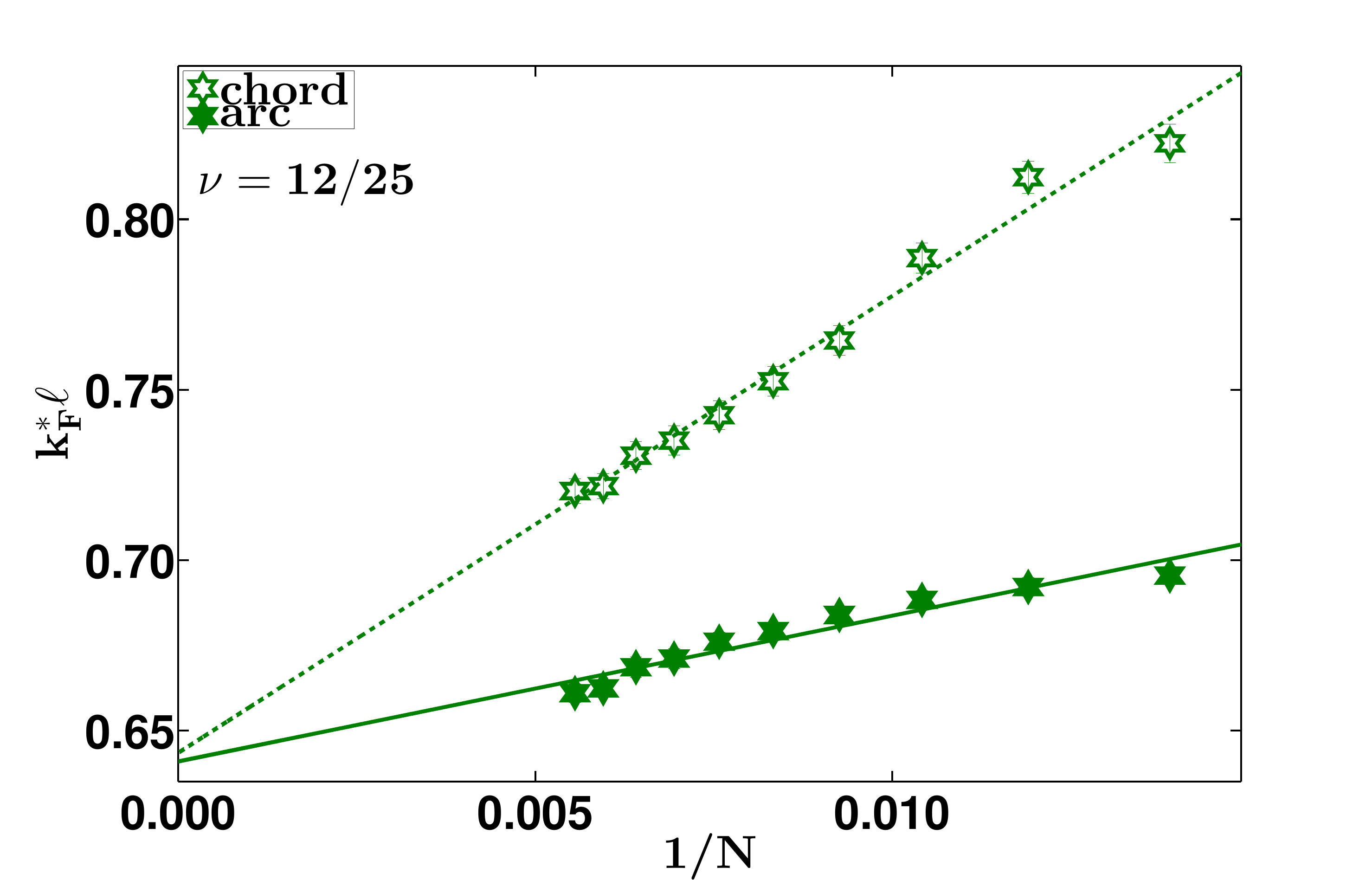}
\includegraphics[width=7.5cm,height=4.5cm]{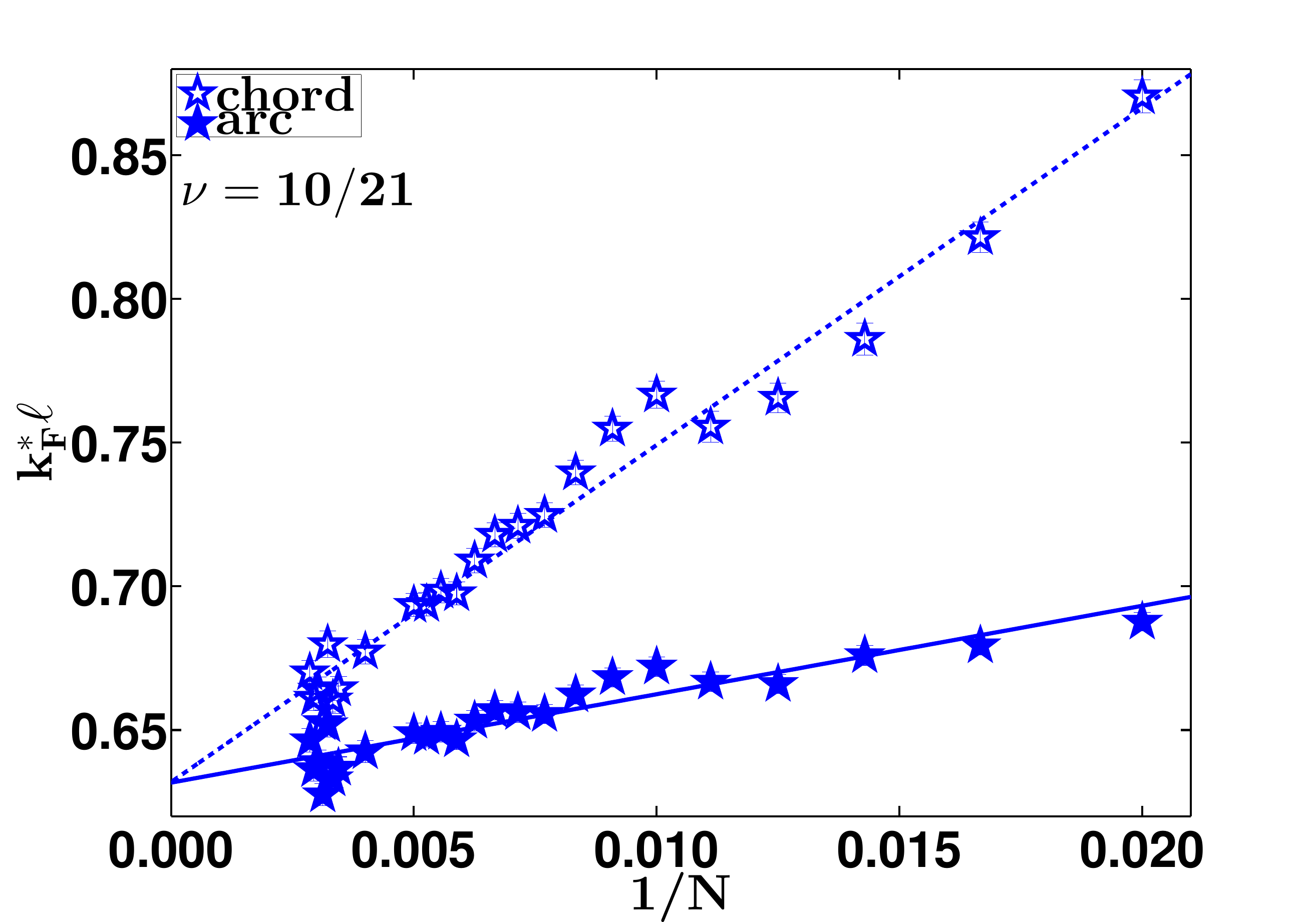} \includegraphics[width=7.5cm,height=4.5cm]{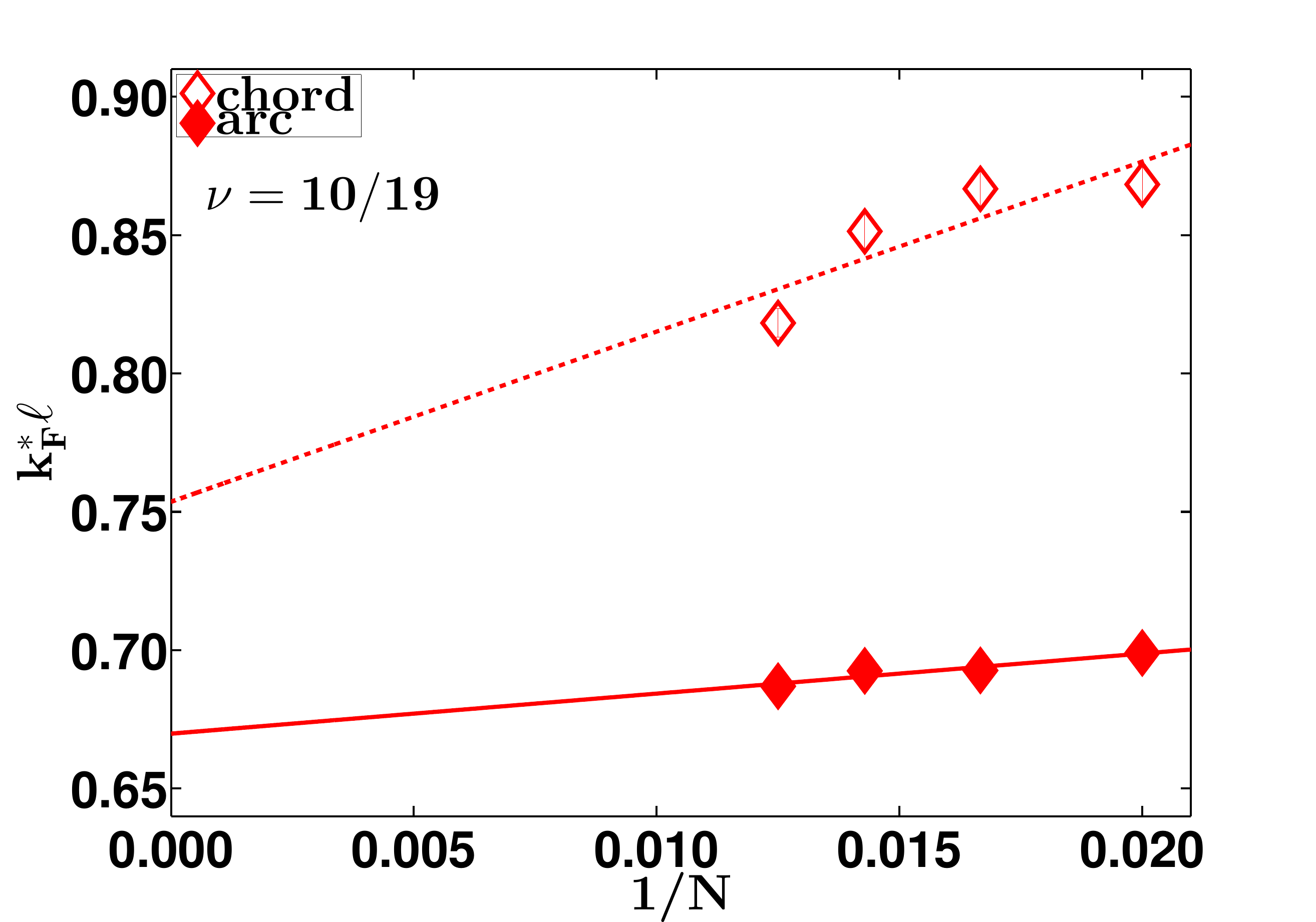} 
\includegraphics[width=7.5cm,height=4.5cm]{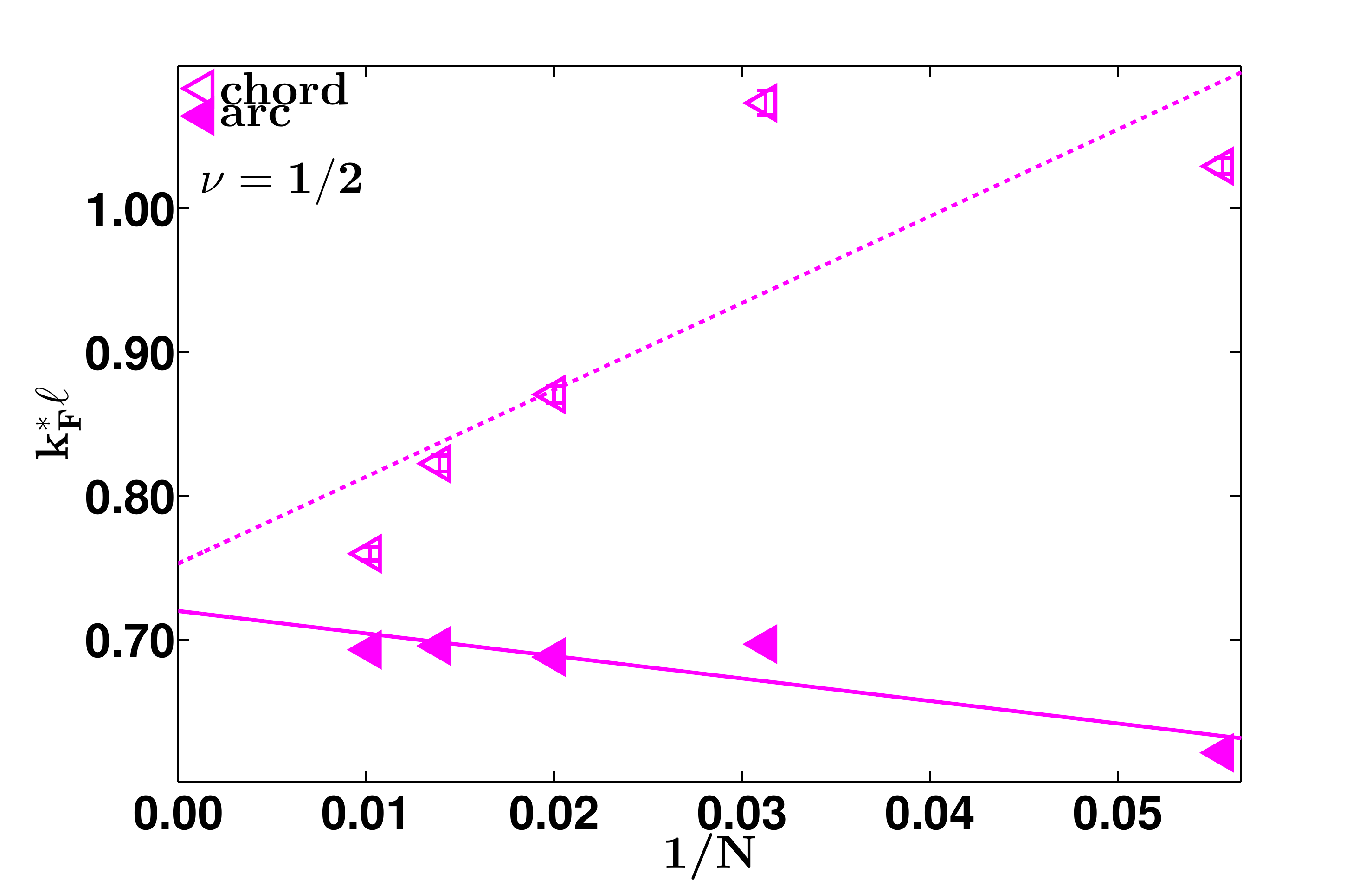} 
\end{center}
\caption{Thermodynamic extrapolation of the Fermi wave vector $k_{\rm F}^*\ell$ for the projected spin-singlet Jain wave function at various filling
factors along the sequence $n/(2n\pm1)$ and the $\nu=1/2$ CF Fermi sea (bottom-most panel). The empty (filled) symbols correspond to the values obtained from the chord (arc) distance on the sphere and the dashed (thick) lines show linear fits to these values as a function of 1/N.}
\label{kF_extrap}
\end{figure*}

\begin{figure*}[t]
\begin{center}
\includegraphics[width=7.5cm,height=4.5cm]{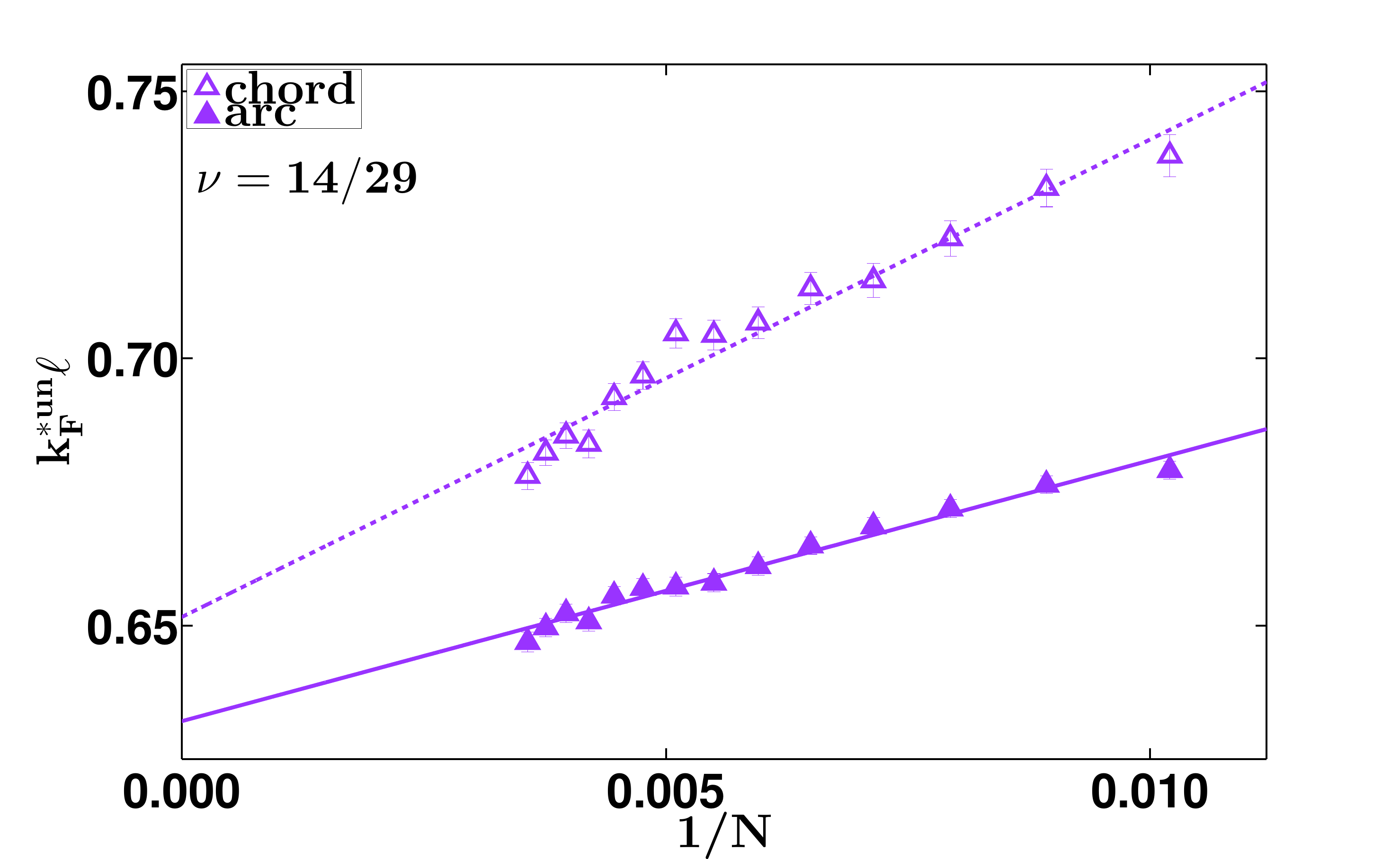} \includegraphics[width=7.5cm,height=4.5cm]{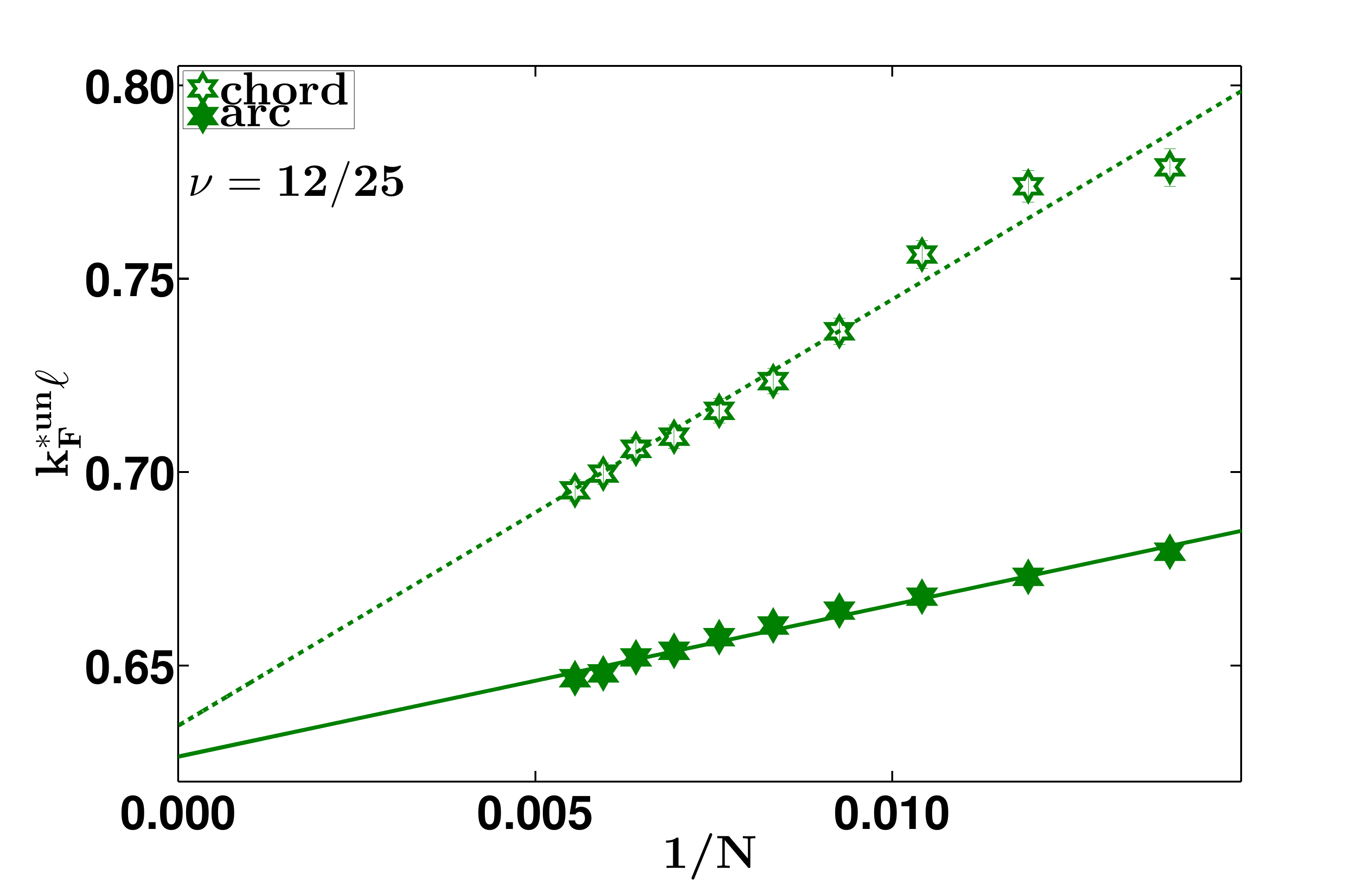}
\includegraphics[width=7.5cm,height=4.5cm]{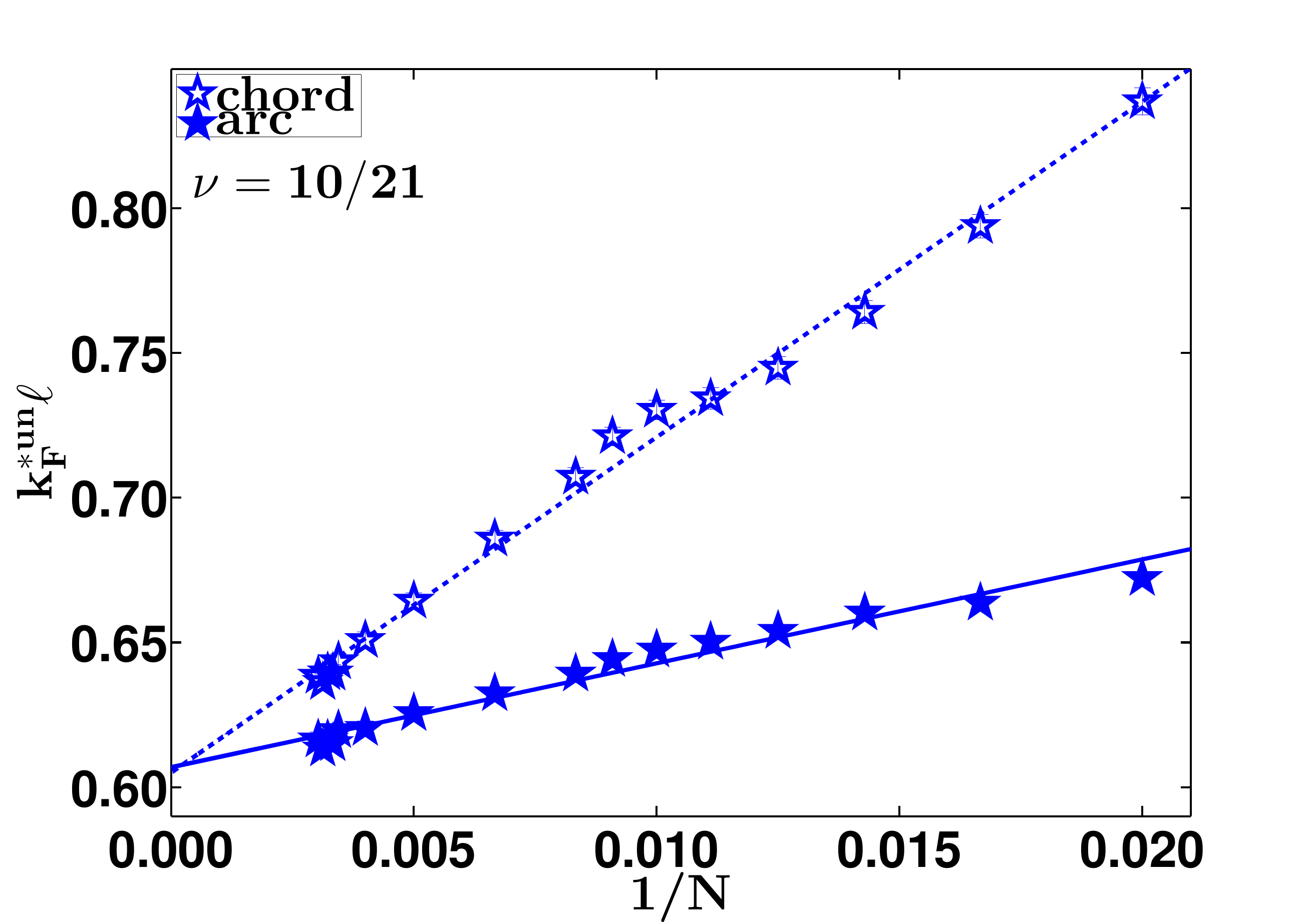} \includegraphics[width=7.5cm,height=4.5cm]{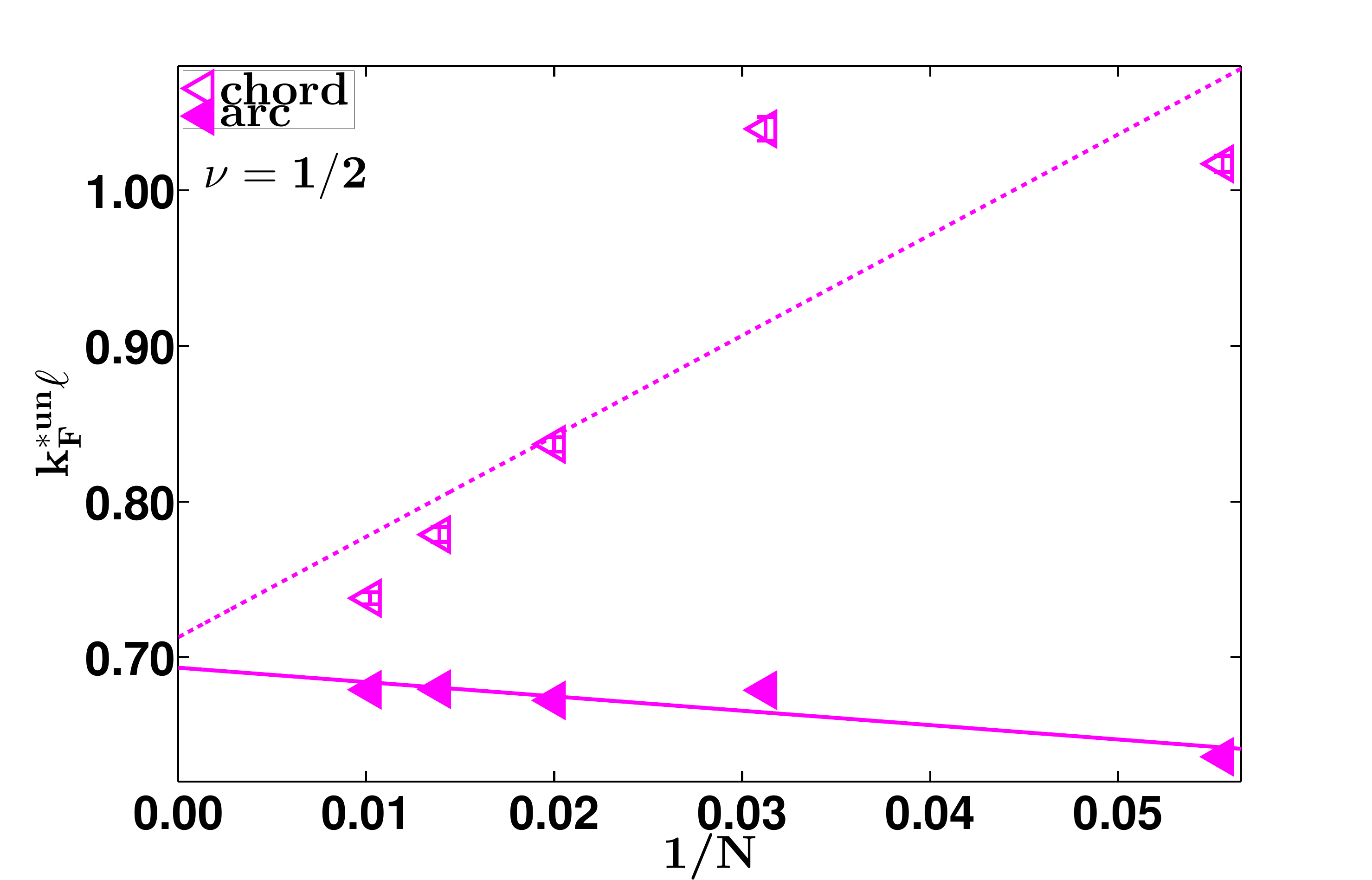}
\end{center}
\caption{Same as Fig.~\ref{kF_extrap} but for the thermodynamic extrapolation of the Fermi wave vector $k_{\rm F}^{*\rm un}\ell$ for the unprojected Jain wave function defined in Eq.~(\ref{eq_unproj_CF}) at various filling factors along the sequence $n/(2n+1)$.}
\label{kFun_extrap}
\end{figure*}

In this appendix we show the thermodynamic extrapolation of the Fermi wave vector obtained from finite size systems in the spherical geometry for spin singlet states. The extrapolations of the Fermi wave vectors for the projected and unprojected states are shown in Fig.~\ref{kF_extrap} and \ref{kFun_extrap} respectively. The thermodynamic values of the Fermi wave vector obtained from these extrapolations are shown in Fig.~\ref{kF}(b) and Fig.~\ref{kFun}(b) where the range shown is obtained from linear fitting in $1/N$ of the Fermi wave vector obtained from the arc and chord distance results. 

Some remarks regarding the extrapolation are in order. At filling factors where we have a large number of systems, the thermodynamic values of the Fermi wave vector given by a linear fit to the arc and chord distance data approach one another. We have also tested that a quadratic fit gives a value within 0.05 of the linear fit. At filling factors $n/(2n+1)$ we have considered very large systems, so we expect corrections to be small. For $\nu=1/2$ and $\nu=n/(2n-1)$ we only have a few systems and, given the scatter, a quadratic fit is not appropriate for them.

\section{Ground state Coulomb interaction energies of spin-singlet states}
\label{app:gs_Coulomb_energies}

For completeness, we list the Coulomb energies of the Jain wave functions for various spin-singlet states along the sequence $n/(2n\pm 1)$. Our calculations are performed in the spherical geometry\cite{Haldane83} using standard Monte Carlo methods, assuming zero thickness and no LL mixing. The density for a finite system in the spherical geometry depends on the number of electrons $N$ and is different from its thermodynamic value. To eliminate this effect we use the so-called ``density-corrected'' energy\cite{Jain07} $E_{N}^{'}=(\frac{2Q\nu}{N})^{1/2}E_{N}$ for extrapolation to the thermodynamic limit $N\rightarrow\infty$. All energies quoted here are the per particle density-corrected energies $E_{N}^{'}/N$.

\begin{figure*}
\begin{center}
\includegraphics[width=5.5cm,height=3.5cm]{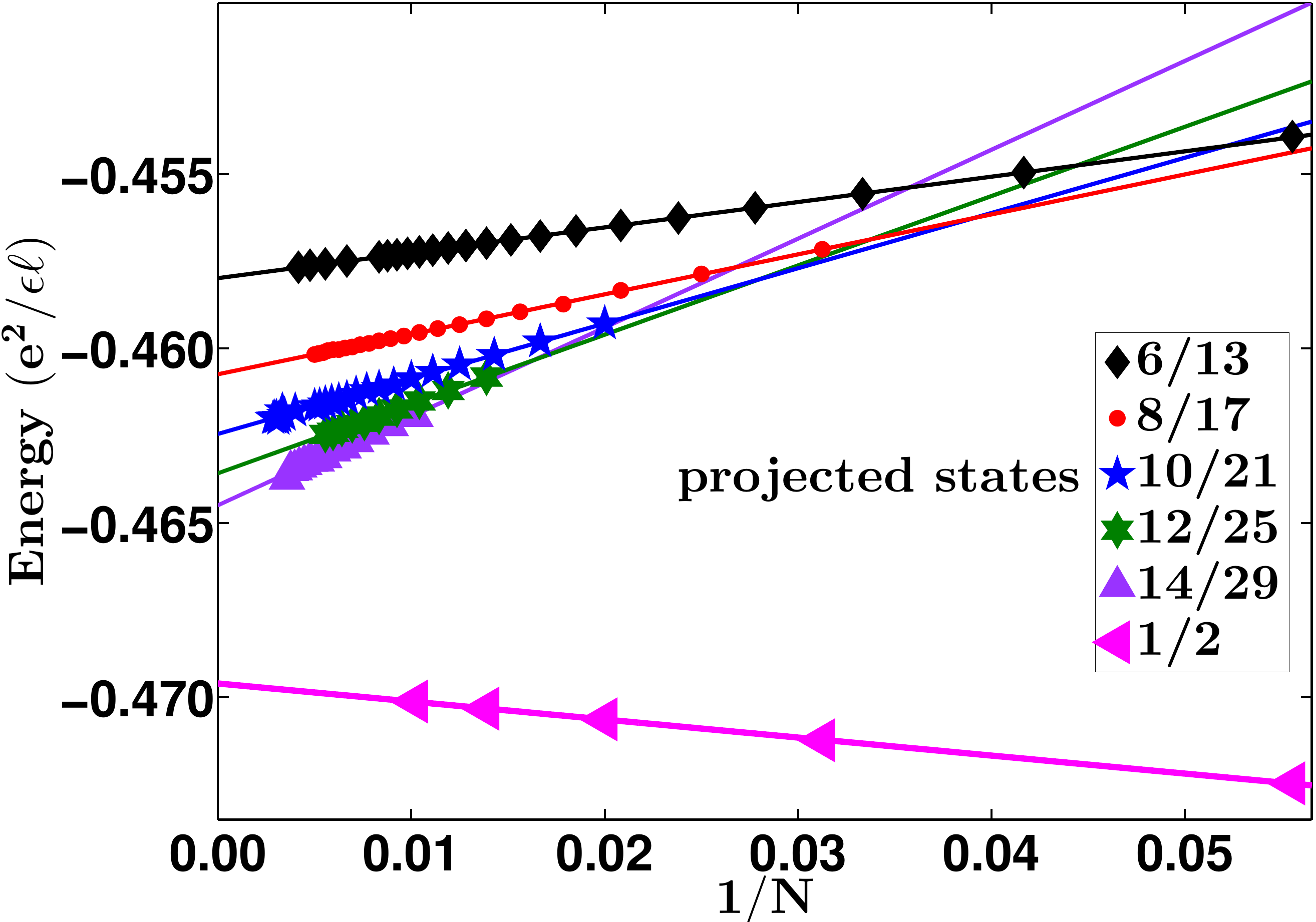} 
\includegraphics[width=5.5cm,height=3.5cm]{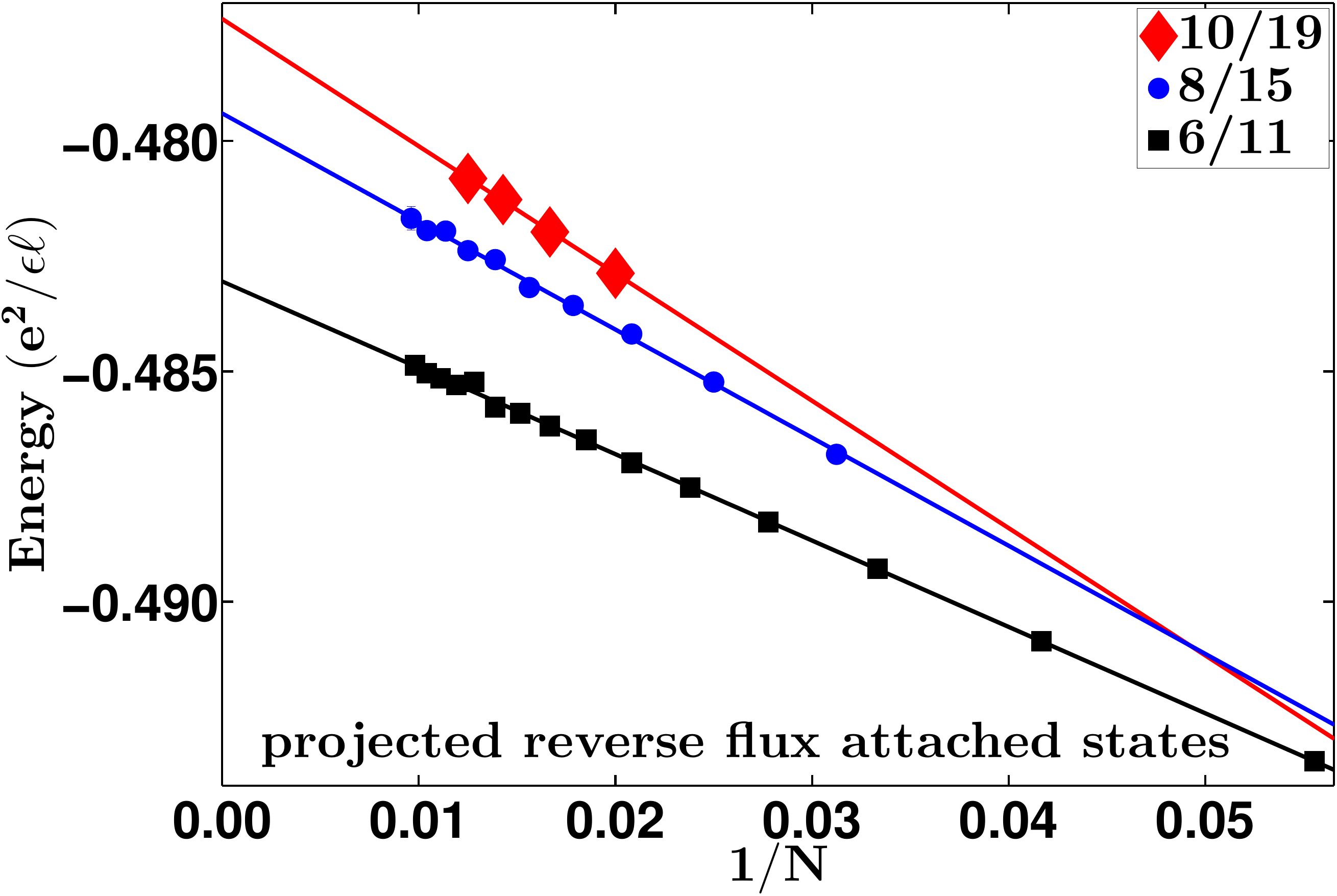}
\includegraphics[width=5.5cm,height=3.5cm]{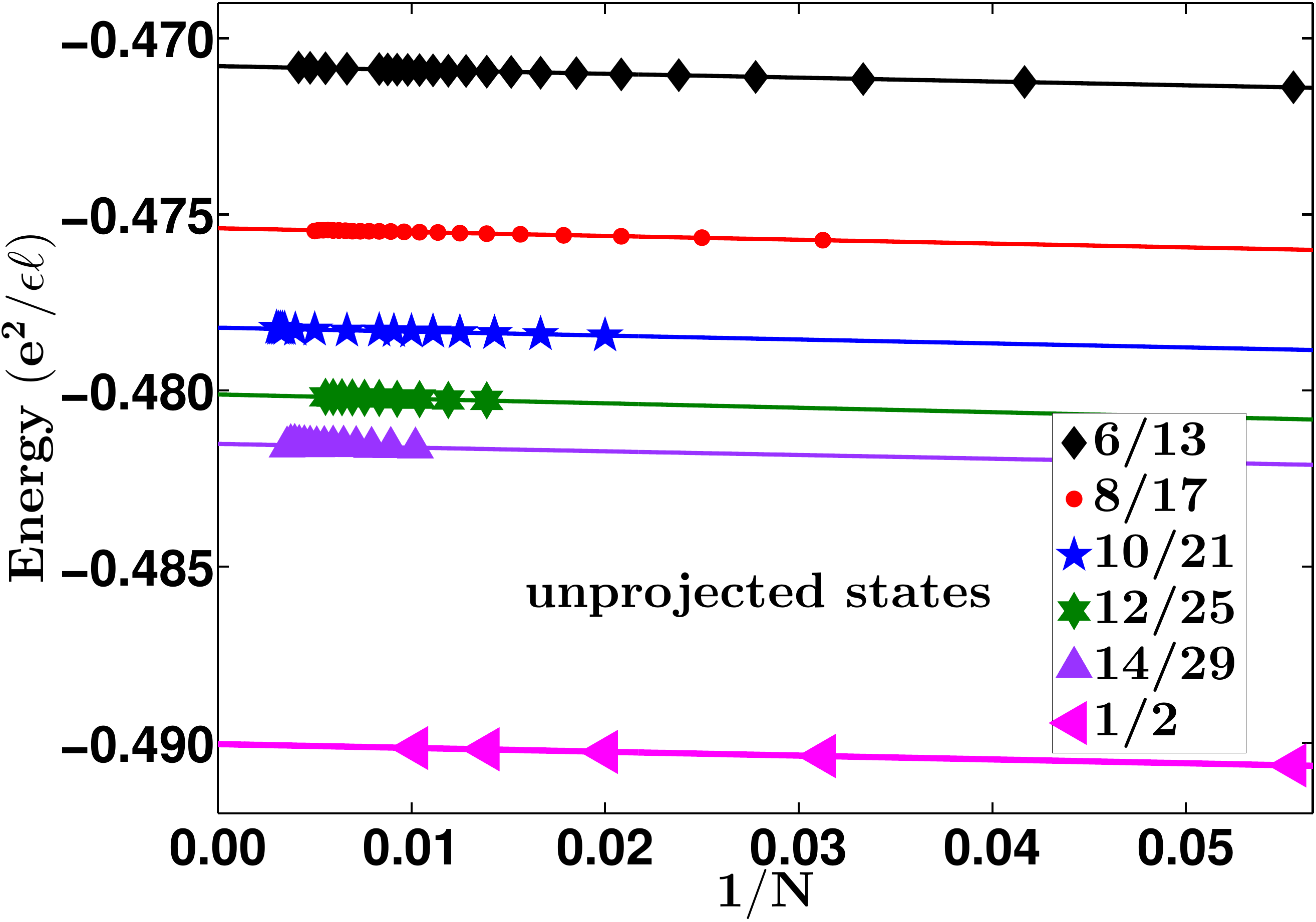}
\end{center}
\caption{(Color online) Thermodynamic extrapolation of the Coulomb ground state energies in the spherical geometry for projected (left and center panels) and unprojected (right panel) spin-singlet states in the sequence $n/(2n\pm 1)$ with even $n$. The extrapolated energies are listed in Table \ref{tab:ss_Coulomb_energies}.}
\label{fig:ss_Coulomb_energies}
\end{figure*}

Fig.~\ref{fig:ss_Coulomb_energies} shows the thermodynamic extrapolation of the projected and unprojected ground state Coulomb energies for the various spin-singlet FQH states along the sequence $n/(2n\pm1)$ as well as for the limiting case of the $\nu=1/2$ composite fermion Fermi sea. In Table \ref{tab:ss_Coulomb_energies} we list out the extrapolated values of the ground state energies. (The kinetic energies of the unprojected states are not included.) As anticipated, the Coulomb energies of the unprojected states are slightly smaller than those of the projected states\cite{Kamilla97b}, because the unprojected states have better correlations at short distance. We note that the ground state energies for the unprojected states along the sequence $n/(2n-1)$ is related to the energies of $n/(2n+1)$ FQH states. This is because $|\Psi^{\rm un}|^2$ evaluated on the unit sphere for the unprojected states at filling factors $\nu=n/(2n-1)$ and $\nu=n/(2n+1)$ is the same for a given $N$. The Coulomb energies scale inversely with the radius of the sphere ($\sqrt{Q}\ell$), thus we have the relation:
\be
{E^{\rm un}_{n\over 2n-1} \over E^{\rm un}_{n \over 2n+1}} = \sqrt{ {N-1+Q^{*} \over N-1-Q^{*}}} ,~Q^{*}= {2N-n^2 \over 4n}
\ee
where $Q^{*}$ is the effective magnetic flux seen by the CFs. This in the thermodynamic limit implies:
\be
E^{\rm un}_{n\over 2n-1} = \left({2n +1 \over 2n-1}\right)^{1/2} E^{\rm un}_{n \over 2n+1}
\label{Erelation_un}
\ee

\begin{table}[ht]
\begin{center}
\begin{tabular}{|c|c|c|}
\hline
\multicolumn{1}{|c|}{filling factor $\nu$} & \multicolumn{2}{|c|}{per particle Coulomb energy $(e^2/\epsilon\ell)$} \\ \hline
					   & projected state			& unprojected state 	 \\ \hline
6/13 					   & -0.45798(1)	 	& -0.47080(1)		\\ \hline
8/17 					   & -0.46074(1)	 	& -0.47539(0)		\\ \hline
10/21 					   & -0.46245(2)	 	& -0.47821(0)		\\ \hline
12/25 					   & -0.46358(2)	 	& -0.48011(1)		\\ \hline
14/29 					   & -0.46449(5)	 	& -0.48151(2)		\\ \hline
1/2 					   & -0.46961(1)	 	& -0.49003(0)		\\ \hline
6/11 					   & -0.48304(4)	 	& -0.51181(1)		\\ \hline
8/15 					   & -0.47940(8) 		& −0.50609(0)		\\ \hline
10/19 					   & -0.47735(7) 		& −0.50275(0)		\\ \hline
\end{tabular}
\end{center}
\caption {Coulomb interaction energies (in units of $e^2/\epsilon\ell$) obtained from a thermodynamic extrapolation of results on the spherical geometry for spin-singlet ground states states at $\nu=n/(2n\pm1)$. The energies for projected and unprojected states are obtained from the wave functions given in Eq.~(\ref{eq_proj_CF}) and Eq.~(\ref{eq_unproj_CF}) respectively.} 
\label{tab:ss_Coulomb_energies}
\end{table}

\section{Results on fully polarized states}
\label{app:fully_polarized}

We take this opportunity to also report certain results for fully spin polarized states. We will present better estimates for the Fermi wave vector of fully spin polarized composite fermions than those in Ref.~\cite{Balram15b}, obtained from more extensive calculations. We will also present results for the static structure factor for fully spin polarized FQH sates and its comparison with predictions from field theoretical approaches\cite{Gromov15,Nguyen17,Nguyen17b}.

\subsection{Updating results of Ref.~\cite{Balram15b}}

\begin{figure}[t]
\begin{center}
\includegraphics[width=7.5cm,height=4.5cm]{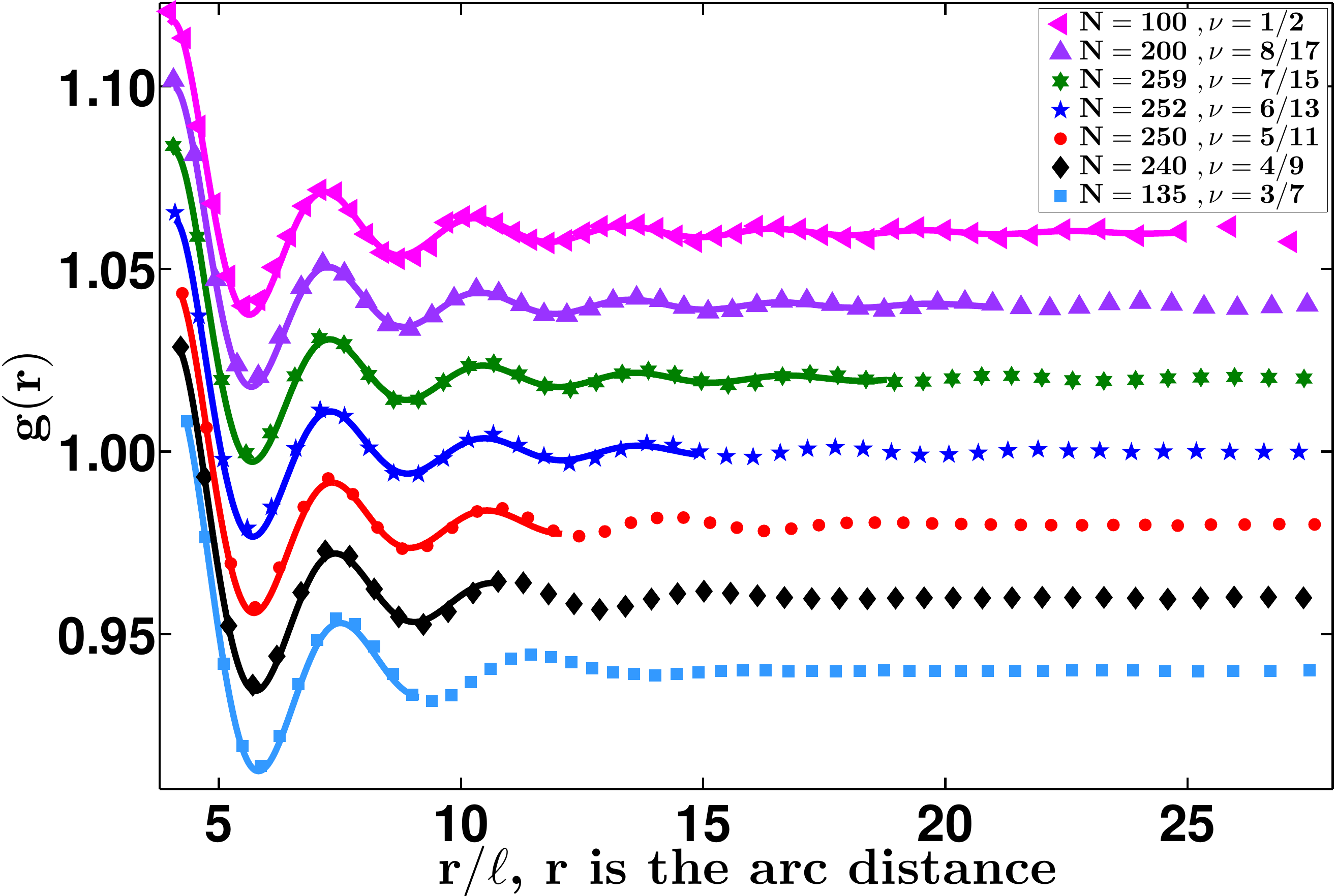}
\includegraphics[width=7.5cm,height=4.5cm]{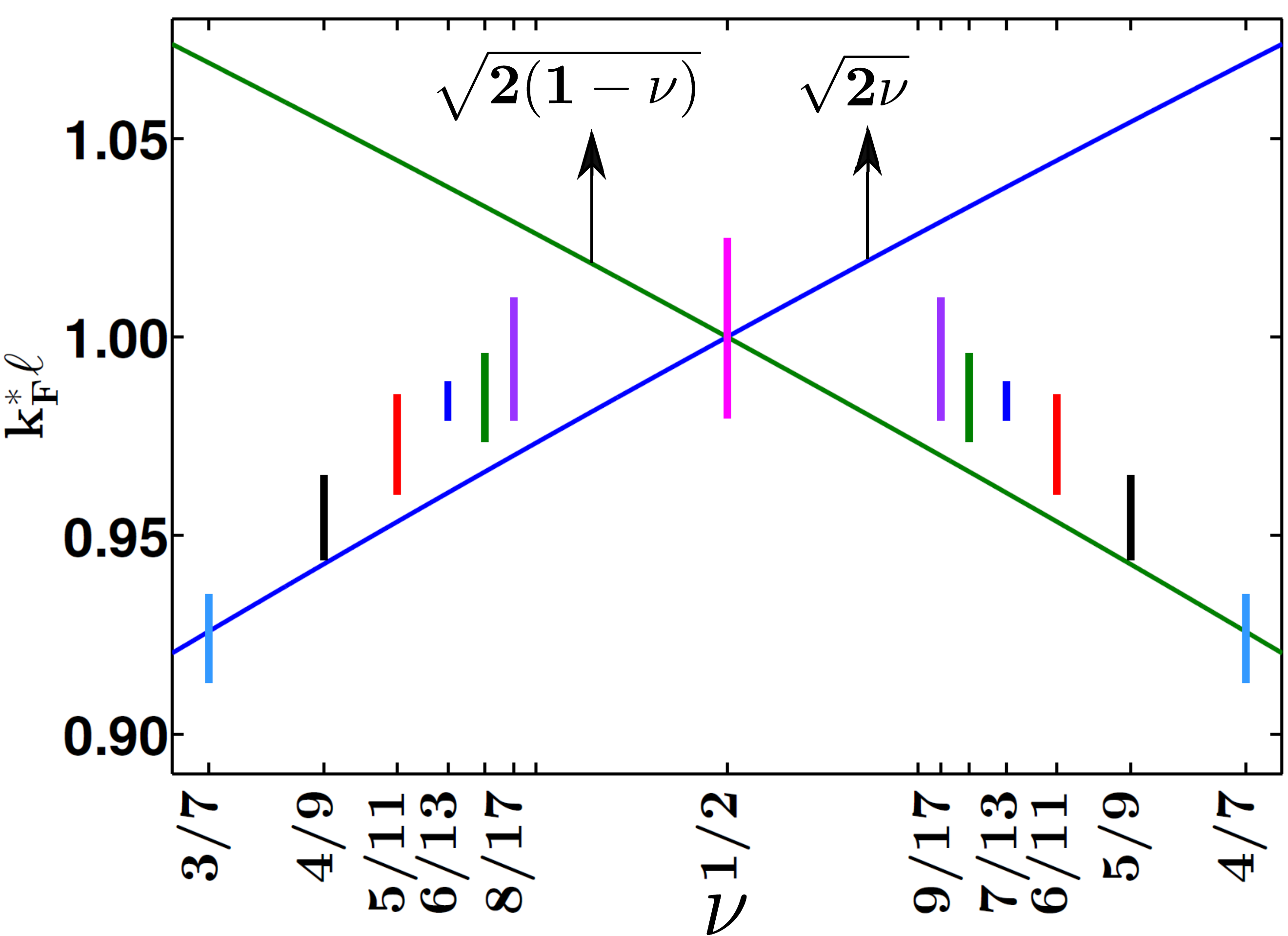}
\end{center}
\caption{(a) The pair correlation function obtained from the projected Jain wave functions of fully spin polarized composite fermions, as a function the arc distance on the sphere. The thick lines are fits to $g(r)=1+A (r\sqrt{4\pi \rho_{\rm e}})^{-\alpha} \sin(2k^*_{\rm F} r+\theta)$ in an intermediate range of $r$ where oscillations are seen. The curves (except for $6/13$) have been shifted up or down by multiples of $0.02$ for ease of viewing. (b) Thermodynamic value of $k_{\rm F}^*\ell$ as a function of $\nu$ obtained from linear and quadratic fits to the arc and chord data. The mean-field values $\sqrt{2\nu}$ (blue) and $\sqrt{2(1-\nu)}$ (green) are shown for reference. }
\label{kF_fp}
\end{figure}

\begin{figure*}[ht]
\begin{center}
\includegraphics[width=5.5cm,height=3.5cm]{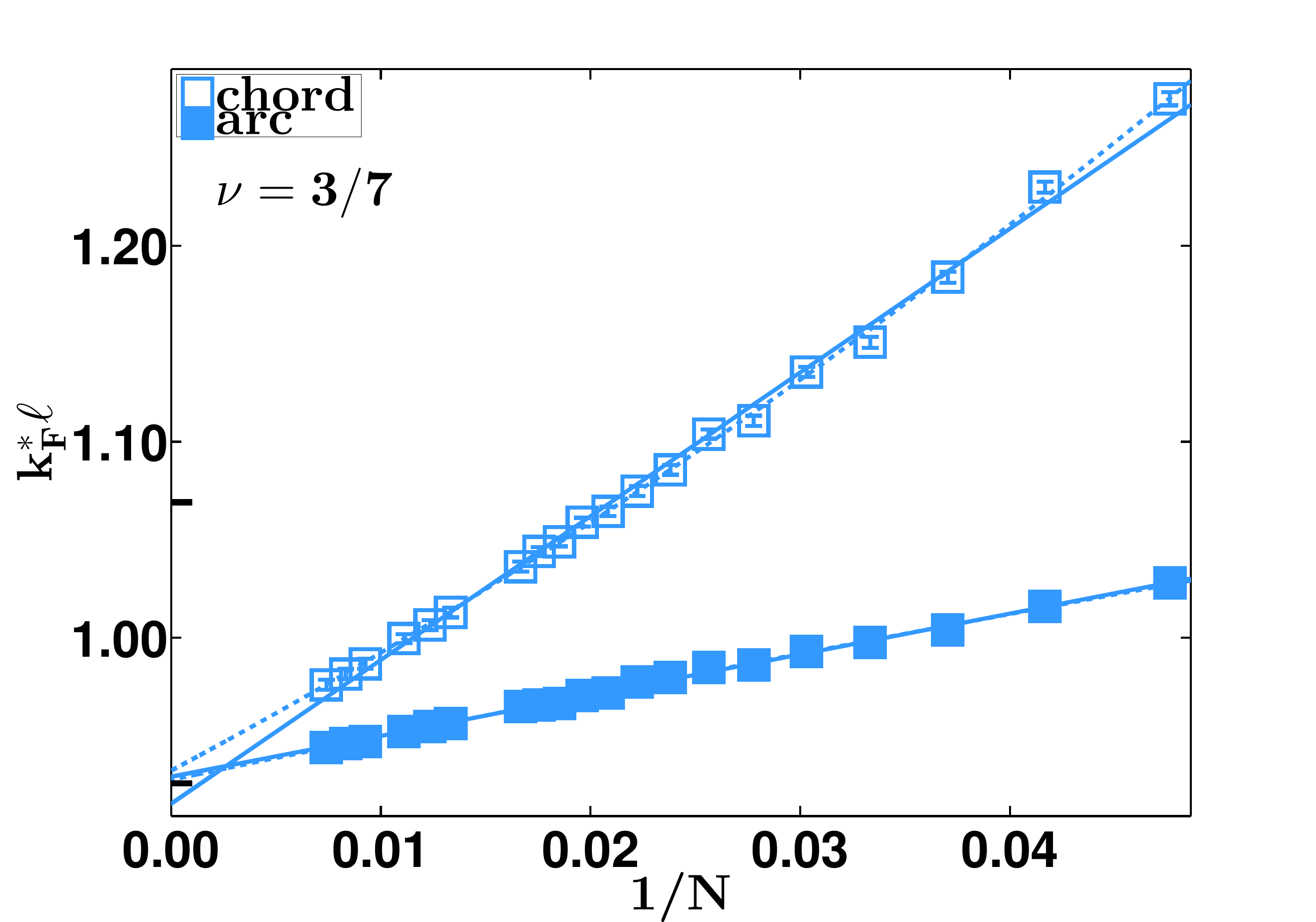} 
\includegraphics[width=5.5cm,height=3.5cm]{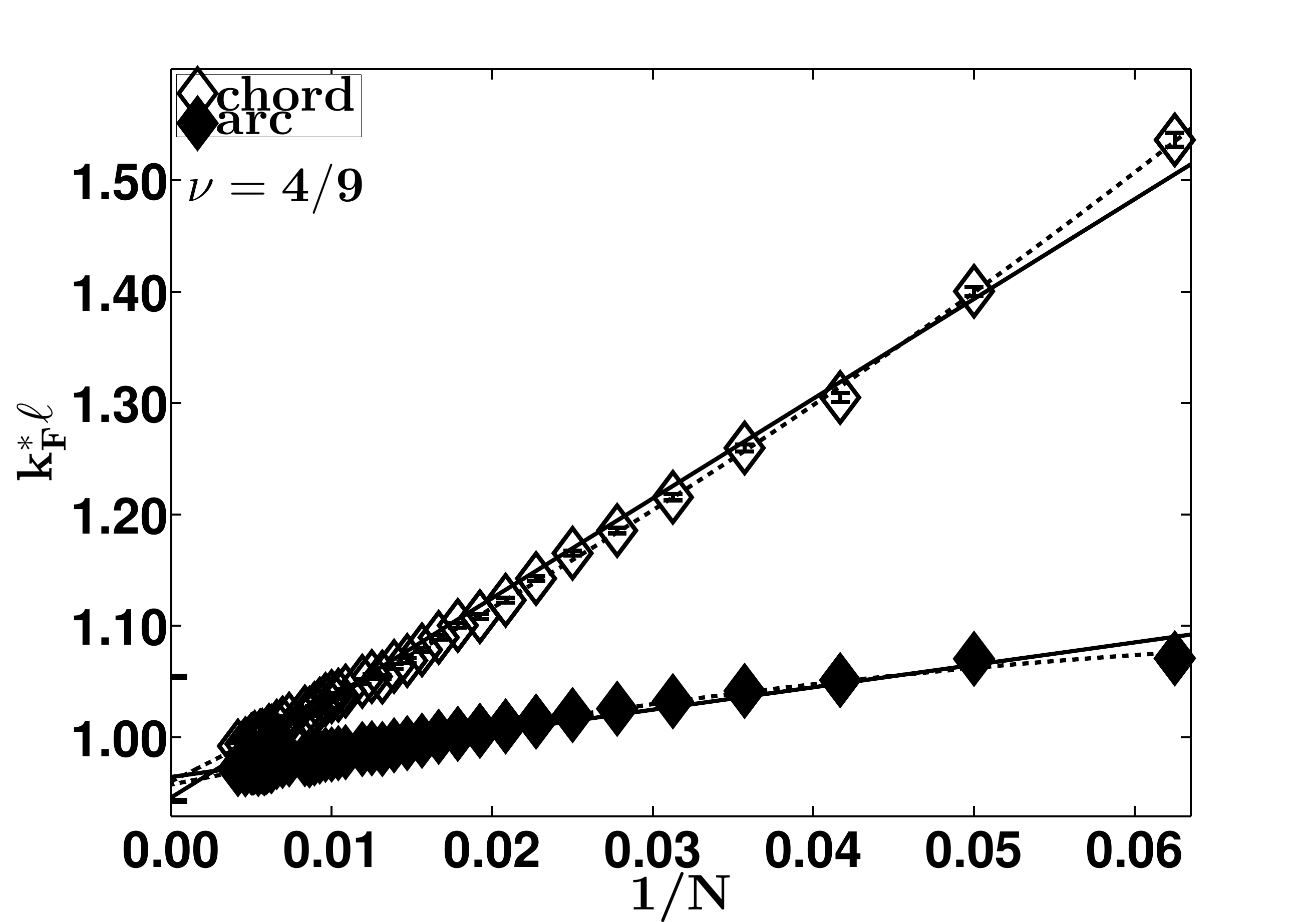} 
\includegraphics[width=5.5cm,height=3.5cm]{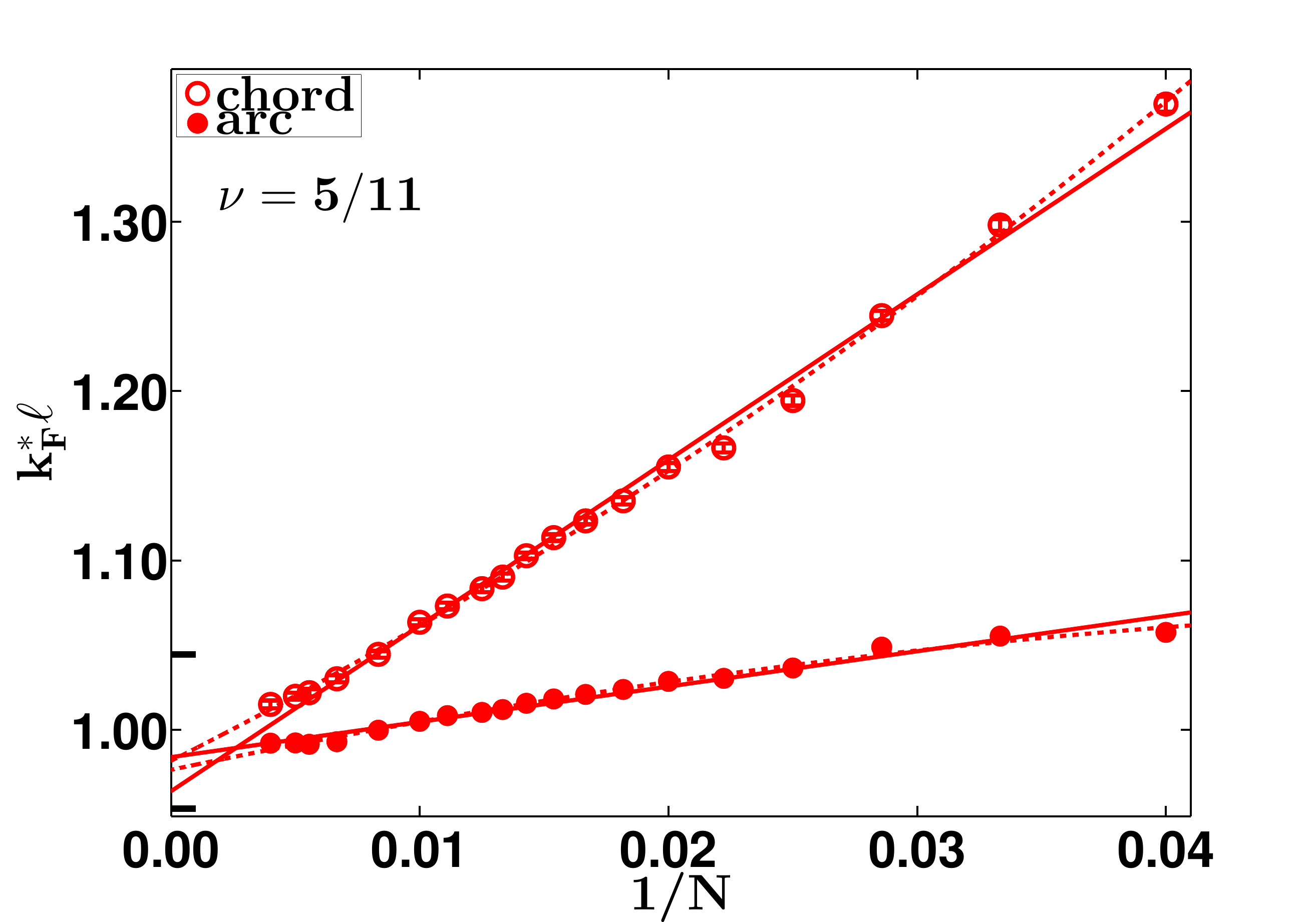} \\
\includegraphics[width=5.5cm,height=3.5cm]{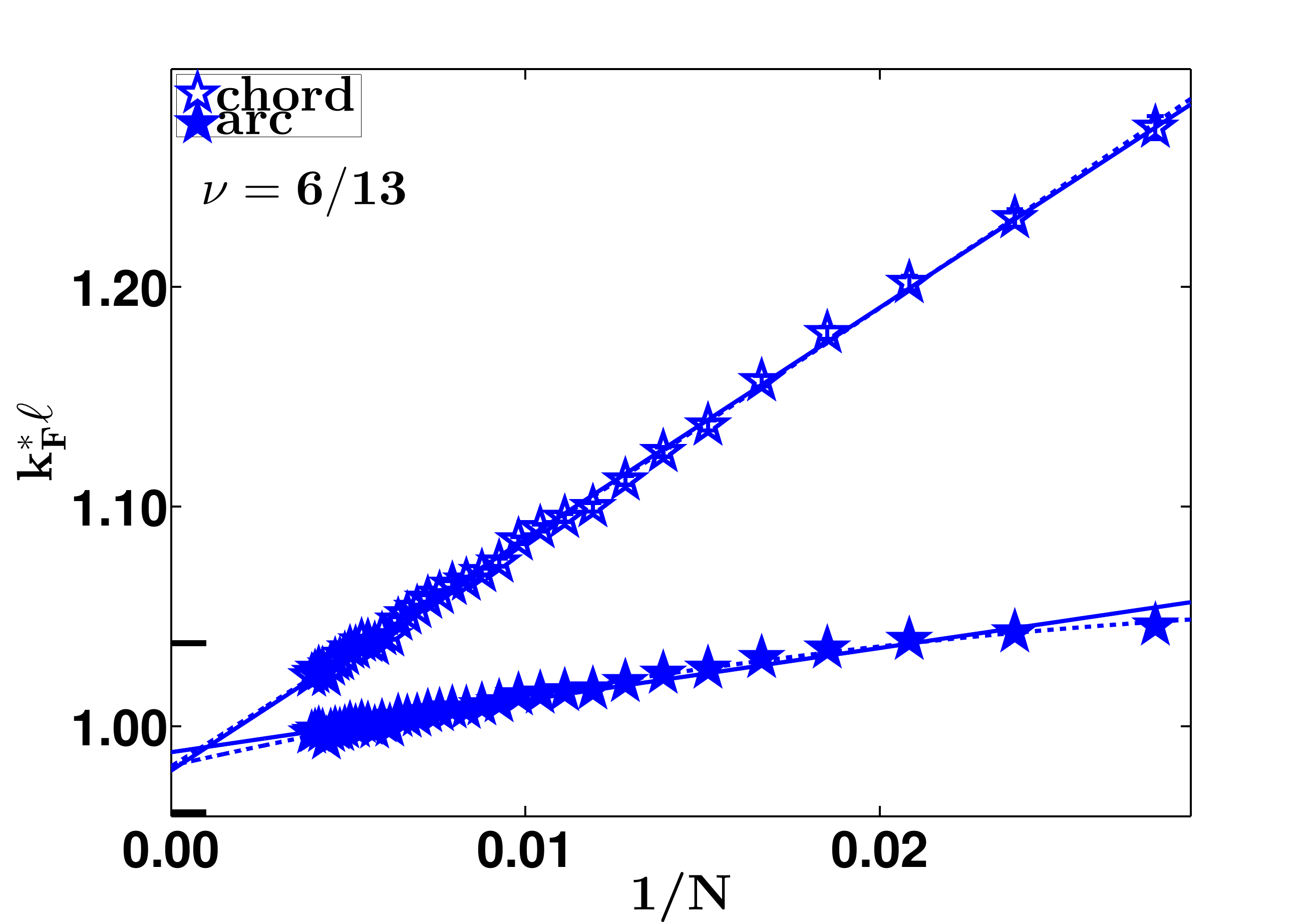} 
\includegraphics[width=5.5cm,height=3.5cm]{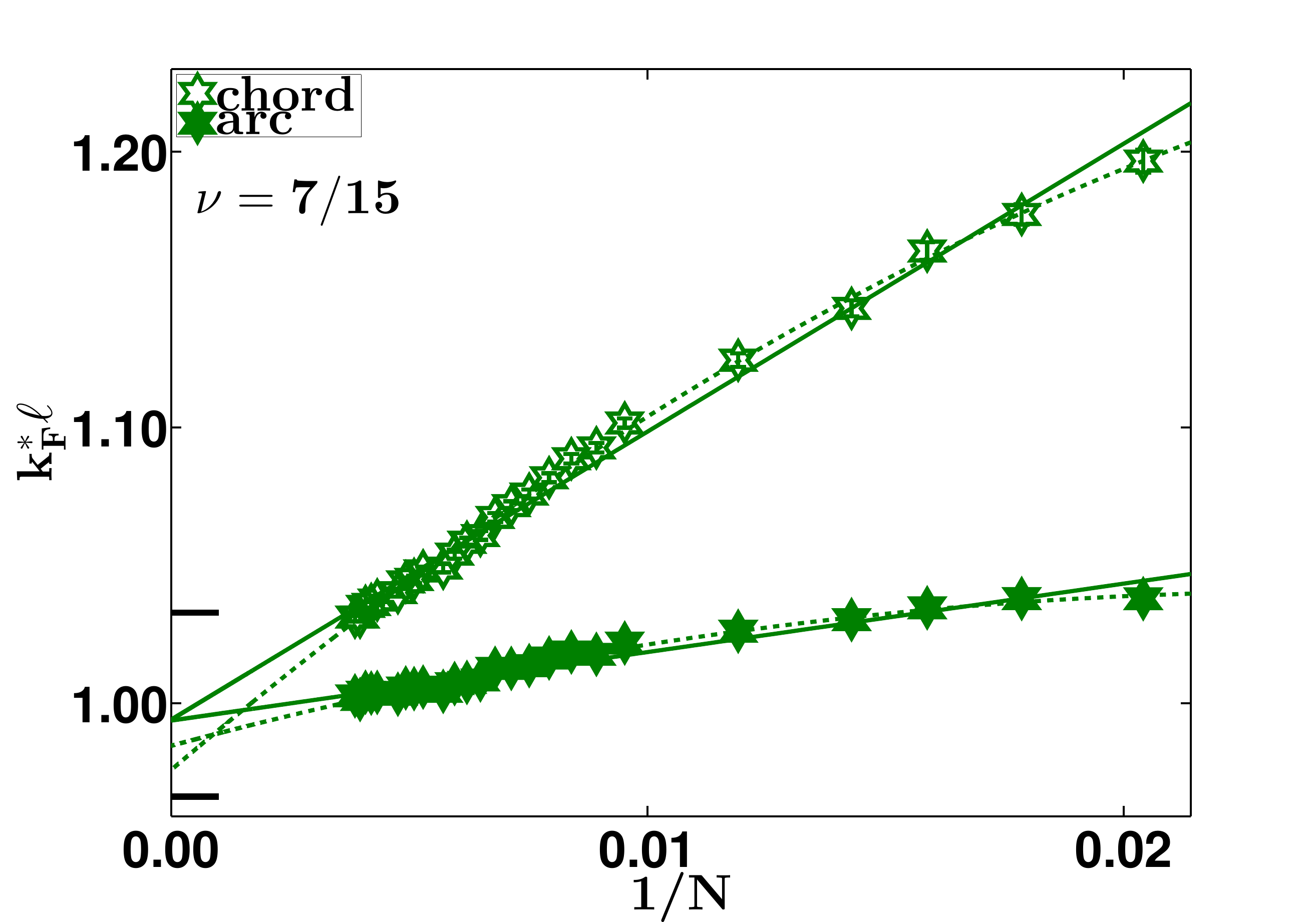} 
\includegraphics[width=5.5cm,height=3.5cm]{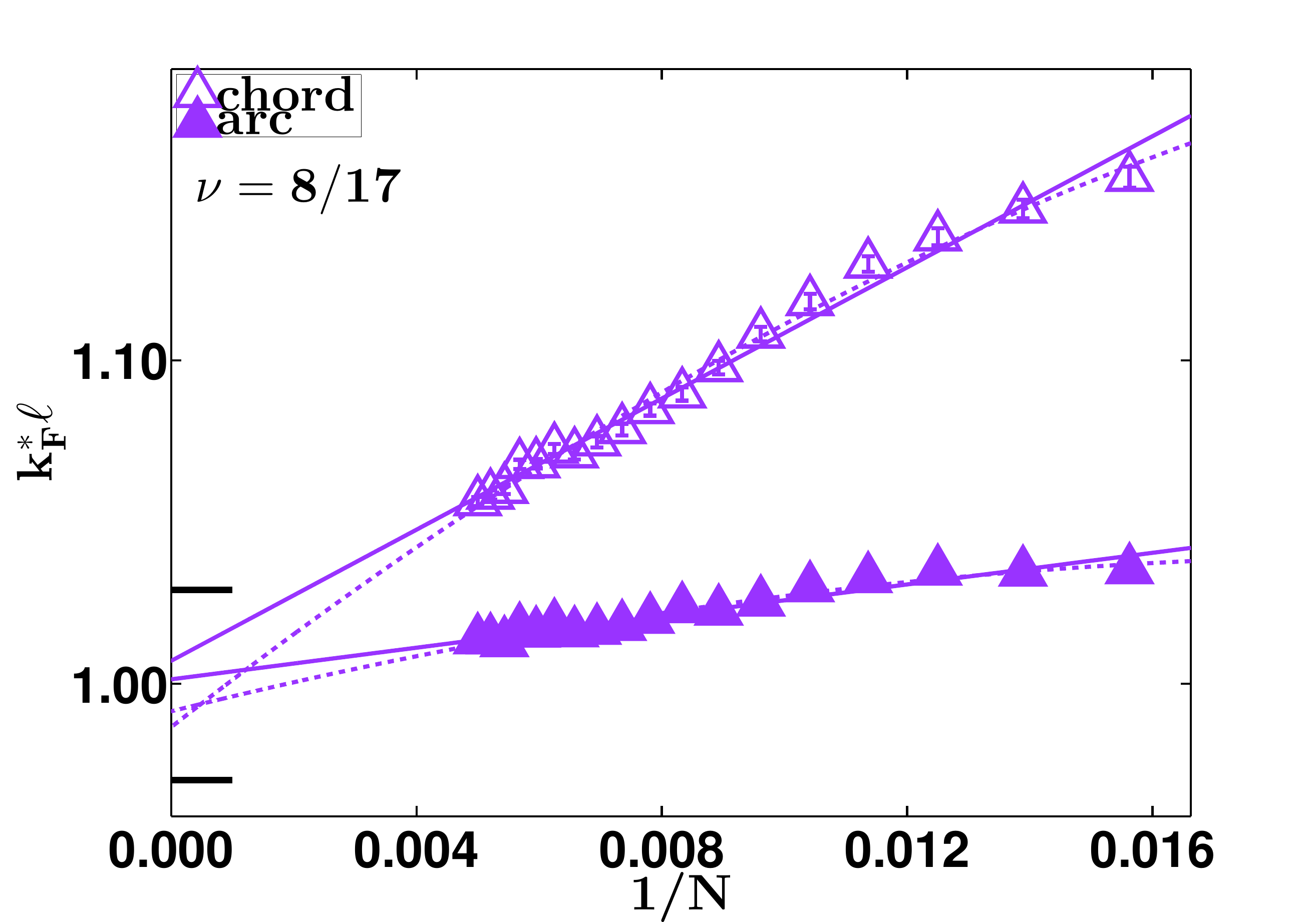} \\
\includegraphics[width=5.5cm,height=3.5cm]{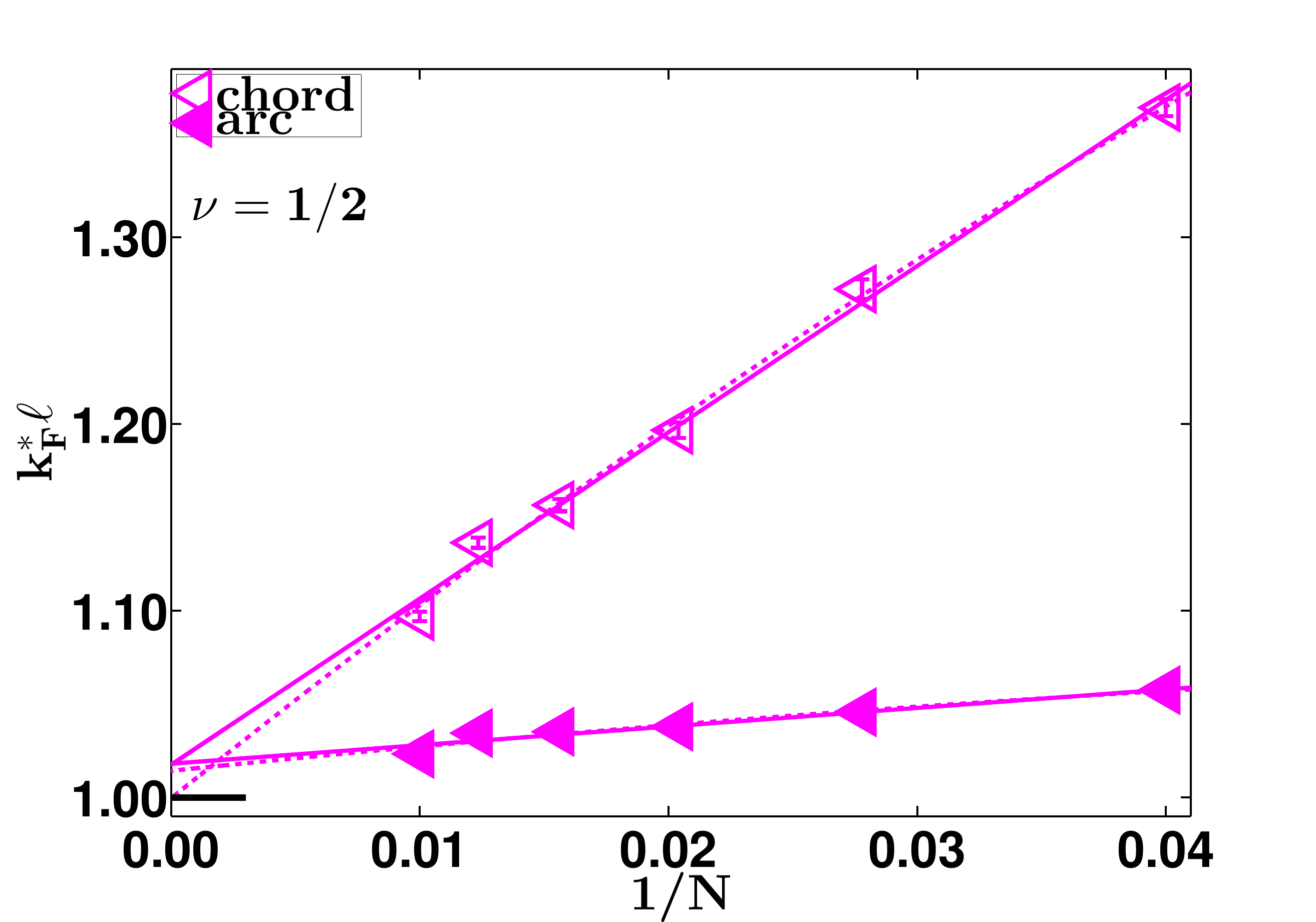} 
\end{center}
\caption{Thermodynamic extrapolation of the Fermi wave vector $k_{\rm F}^*\ell$ for the projected fully polarized Jain wave function at various filling
factors along the sequence $n/(2n+1)$ and the $\nu=1/2$ CF Fermi sea (bottom-most panel). The empty (filled) symbols correspond to the values obtained from the chord (arc) distance on the sphere and the thick and dashed lines show linear and quadratic fits respectively to these values as a function of $1/N$.}
\label{kF_fp_extrap}
\end{figure*}

%Since the publication of Ref.~\cite{Balram15b}, we have done further calculations to obtain better estimates for the Fermi wave vector of fully spin polarized composite fermions, and we take this opportunity to present those results here. 
Since the publication of Ref.~\cite{Balram15b}, we have obtained results for 8/17; we have studied larger systems for states along the sequence $n/(2n+1)$ with $n=3-7$; and we have calculated the pair correlation function for the $\nu=1/2$ state for $N=100$. The updated results are given in Figs.~\ref{kF_fp} and \ref{kF_fp_extrap}. The upper panel of Fig.~\ref{kF_fp} shows the pair correlation function for the largest $N$ and its fitting to $g(r)=1+A (r\sqrt{4\pi \rho_{\rm e}})^{-\alpha} \sin(2k^*_{\rm F} r+\theta)$ to obtain $k^*_{\rm F}$, and Fig.~\ref{kF_fp_extrap} shows thermodynamic extrapolations of $k^*_{\rm F}$ for several fillings. The thermodynamic values of the Fermi wave vector obtained from these extrapolations is shown in the lower panel of Fig.~\ref{kF_fp} where the range shown is obtained from linear and quadratic fitting in $1/N$ of the Fermi wave vector obtained from the chord and arc distance results. These calculations further corroborate the results of Ref.~\cite{Balram15b} in that the Fermi wave vector for Jain states at $\nu=n/(2n+1)$ is close to the value ascertained from the minority carrier rule. Furthermore, we find that at $\nu=1/2$ the Fermi wave vector is nicely consistent with Luttinger's rule.

\begin{figure}[ht]
\begin{center}
\includegraphics[width=8.0cm,height=5.0cm]{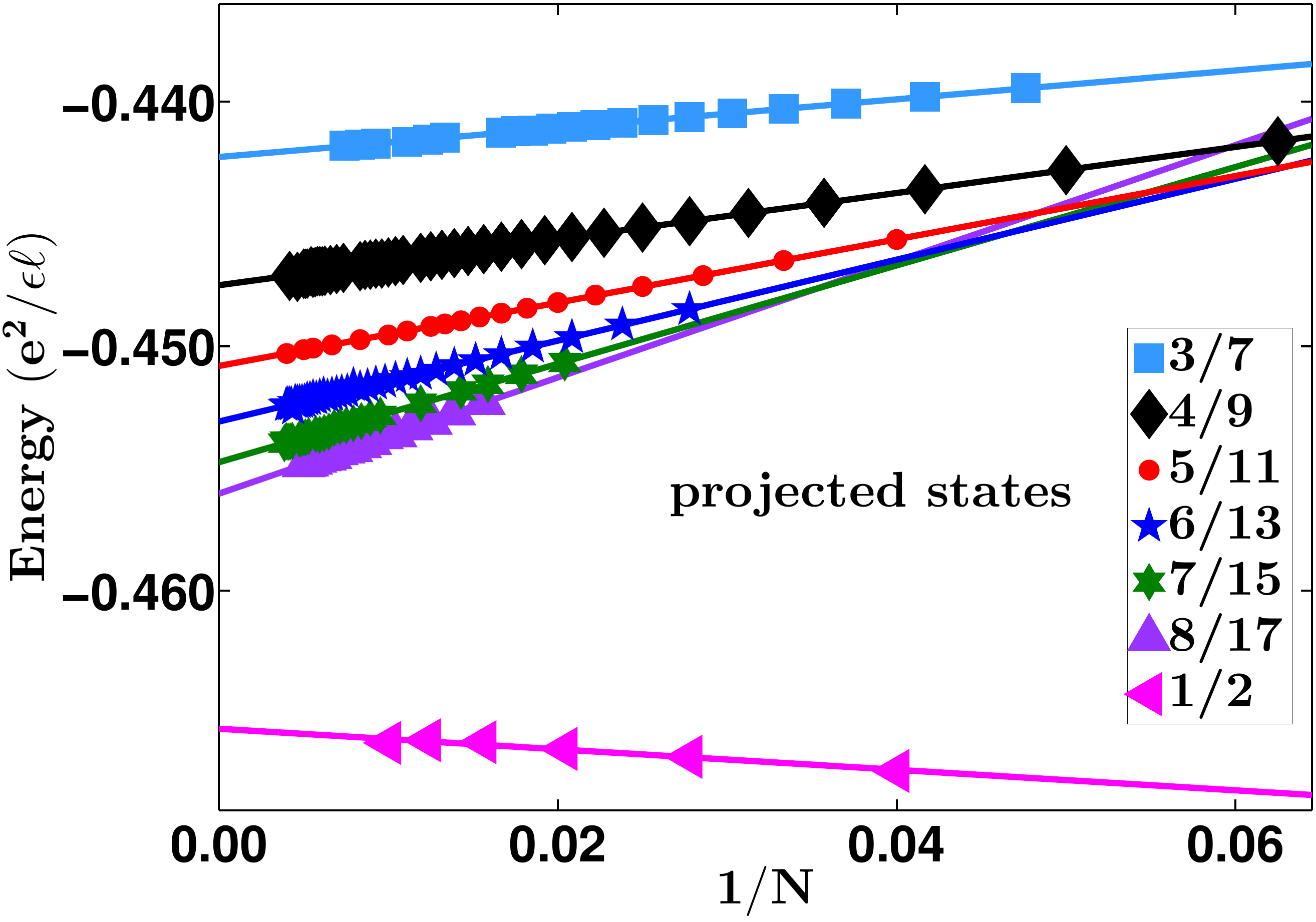} 
\end{center}
\caption{(Color online) Thermodynamic extrapolation of the Coulomb ground state energies in the spherical geometry for the fully polarized Jain states in the sequence $n/(2n+1)$ (see Table \ref{tab:fp_Coulomb_energies} for the extrapolated energies).}
\label{fig:fp_Coulomb_energies}
\end{figure}

For completeness in Fig.~\ref{fig:fp_Coulomb_energies} we show the extrapolation of the Coulomb ground state energies and tabulate them in Table \ref{tab:fp_Coulomb_energies}. 

\begin{table}[ht]
\begin{center}
\begin{tabular}{|c|c|}
\hline
\multicolumn{1}{|c|}{filling factor $\nu$} & \multicolumn{1}{|c|}{per particle Coulomb energy $(e^2/\epsilon\ell)$} \\ \hline
3/7 					   & -0.44226(1)	 	\\ \hline
4/9 					   & -0.44751(1)	 	\\ \hline
5/11 					   & -0.45081(1)	 	\\ \hline
6/13 					   & -0.45309(2)	 	\\ \hline
7/15 					   & -0.45475(1)	 	\\ \hline
8/17 					   & -0.45604(2)	 	\\ \hline
1/2 					   & -0.46566(10)	 	\\ \hline
\end{tabular}
\end{center}
\caption {Coulomb interaction energies (in units of $e^2/\epsilon\ell$) obtained from a thermodynamic extrapolation of results on the spherical geometry for the fully polarized ground states at $\nu=n/(2n+1)$ using the Jain wave function.} 
\label{tab:fp_Coulomb_energies}
\end{table}

\subsection{static structure factor}

We have also considered the static structure factor for the fully spin polarized state and compared it to the predictions by Gromov et al. using a topological approach\cite{Gromov15,Nguyen17} and those based on the Dirac CF description\cite{Nguyen17,Nguyen17b}. We find that our results agree with the predictions in the long wave length limit.

The structure factor $S(\vec{q})$ is defined by the relation\cite{Giuliani08,Kamilla97}:
\begin{equation}
 S(\vec{q})=\frac{\langle \rho_{\vec{q}}\rho_{-\vec{q}} \rangle }{N}-N\delta_{\vec{q},0},~
 \rho_{\vec{q}}=\sum_{j}e^{i\vec{q}.\vec{r}_{j}}
\label{def_Sq}
\end{equation}
where $\langle \cdots \rangle$ denotes the expectation value in the ground state. It is related to the pair-correlation function $g(\vec{r})$ by the Fourier transform\cite{Kamilla97}:
\begin{equation}
 S(\vec{q})-1=\int d^{2}\vec{r}~e^{i\vec{q}.\vec{r}}\rho(\vec{r})[g(\vec{r})-1]
\end{equation}
Considering uniform incompressible states on the plane we get:
\begin{equation}
 S(q)-1= 2\pi\rho \int dr~rJ_{0}(kr)[g(r)-1]
 \label{eq_Sk_gr}
\end{equation}
where $J_{0}(x)$ is the zeroth order Bessel function of the first kind. For the Fermi sea of non-interacting fully polarized electrons at zero magnetic field the static structure factor is given by\cite{Kamilla97}:
\begin{equation}
 S(q)=\begin{cases}
       1, ~q\geq 2k_{F} \\
       \frac{2}{\pi}\left[\frac{q}{2k_{F}}\sqrt{1-\left(\frac{q}{2k_{F}}\right)^2}+\arcsin \left( \frac{q}{2k_{F}} \right) \right],~q< 2k_{F} \\	
      \end{cases}
\end{equation}
which starts at zero for $q=0$ and increases monotonically till it attains its maximum value of unity at $2k_{\rm F}$ and then stays there for $q>2k_{\rm F}$. At $q=2k_{\rm F}$, $S(q)$ as well as its first derivative are both continuous.

Using the pair-correlation functions calculated in the spherical geometry we can get the structure factor by numerically evaluating the integral given in Eq.~(\ref{eq_Sk_gr}). This is a valid approach since the systems considered in this work are large and curvature effects are negiligible for them, thereby allowing us to use the Fourier transform on the plane. One can also directly use the Fourier transform on the sphere and we have checked that these give similar results. 

\begin{figure}[ht]
\begin{center}
\includegraphics[width=0.23\textwidth]{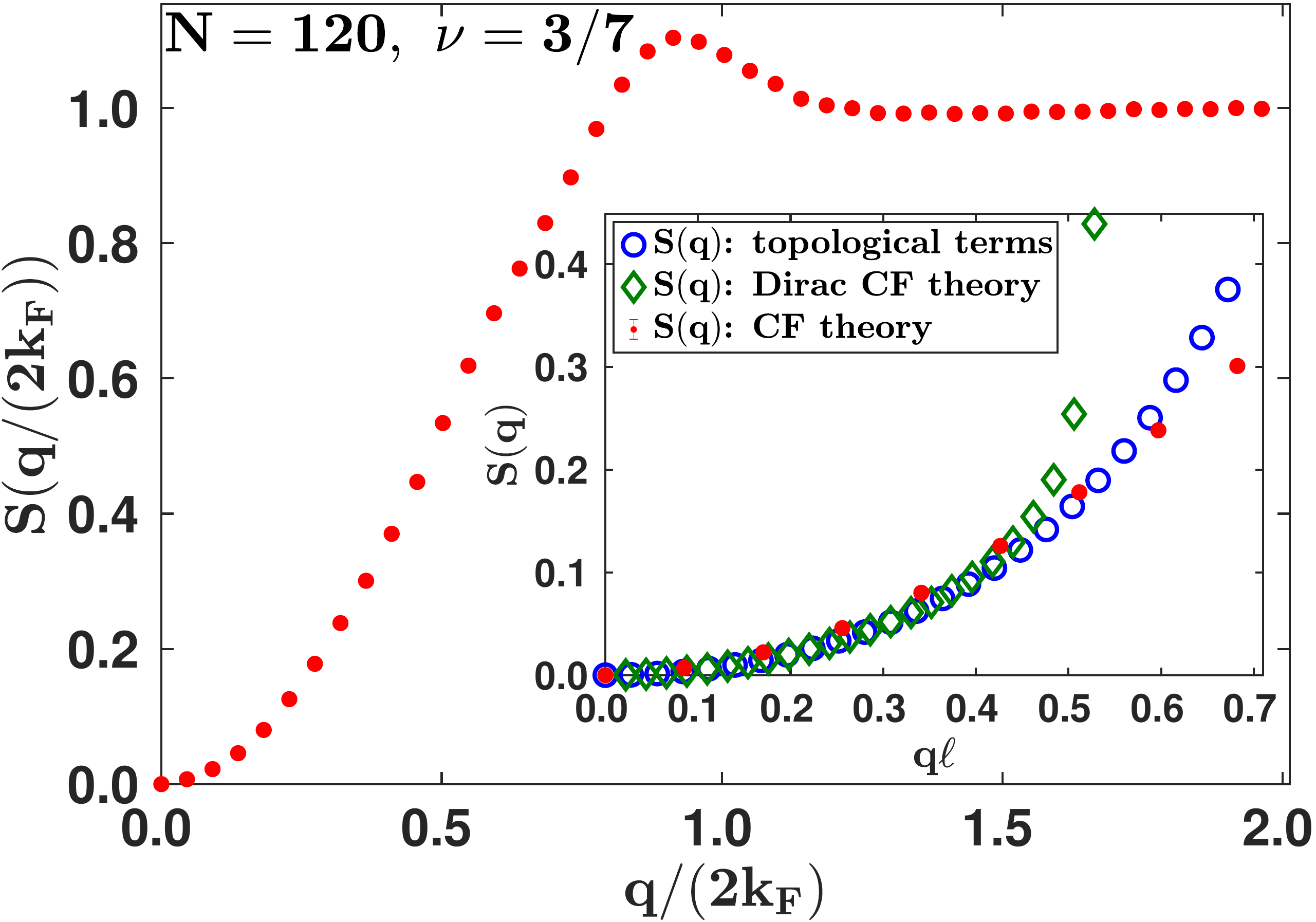}
\includegraphics[width=0.23\textwidth]{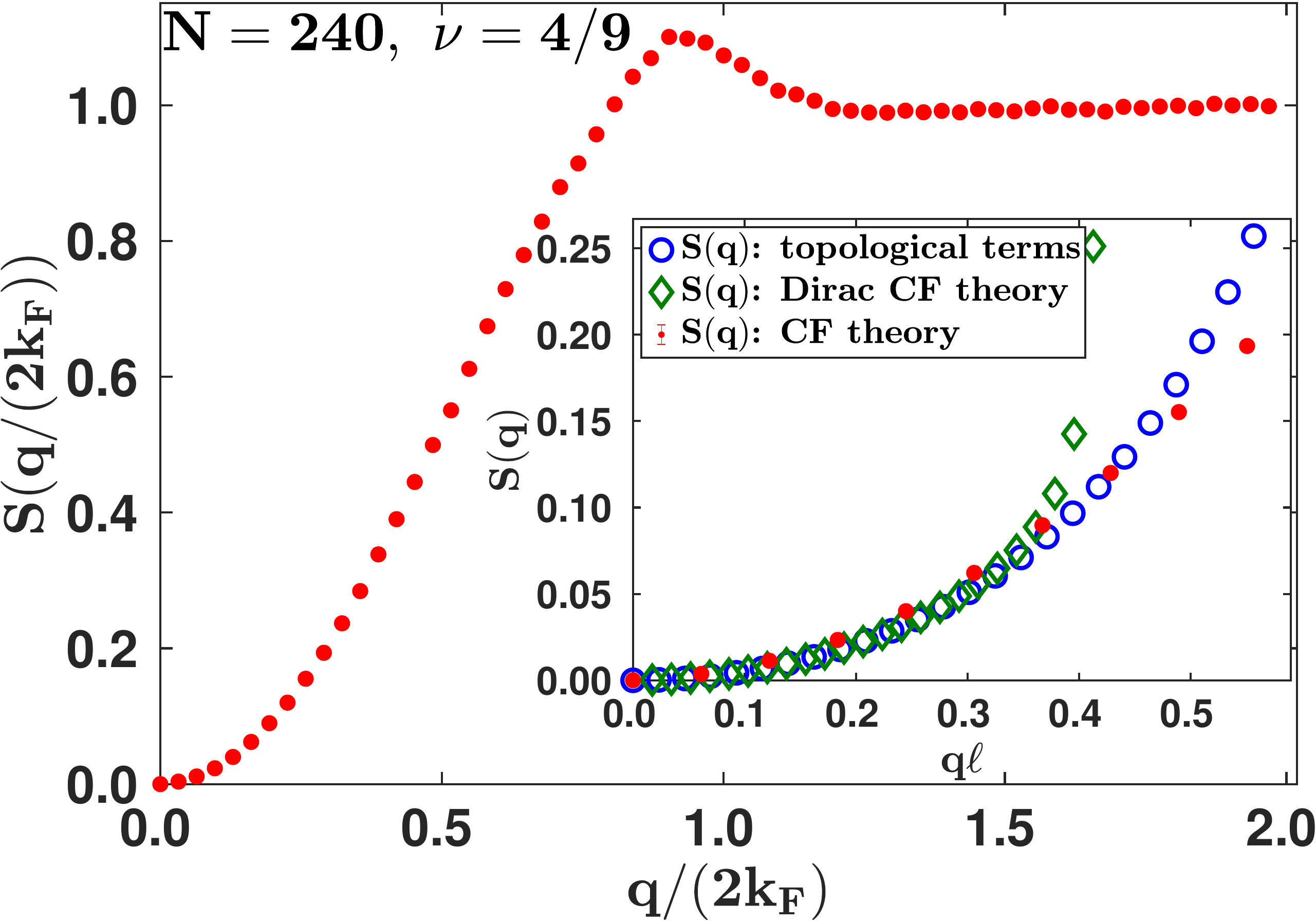} \\
\includegraphics[width=0.23\textwidth]{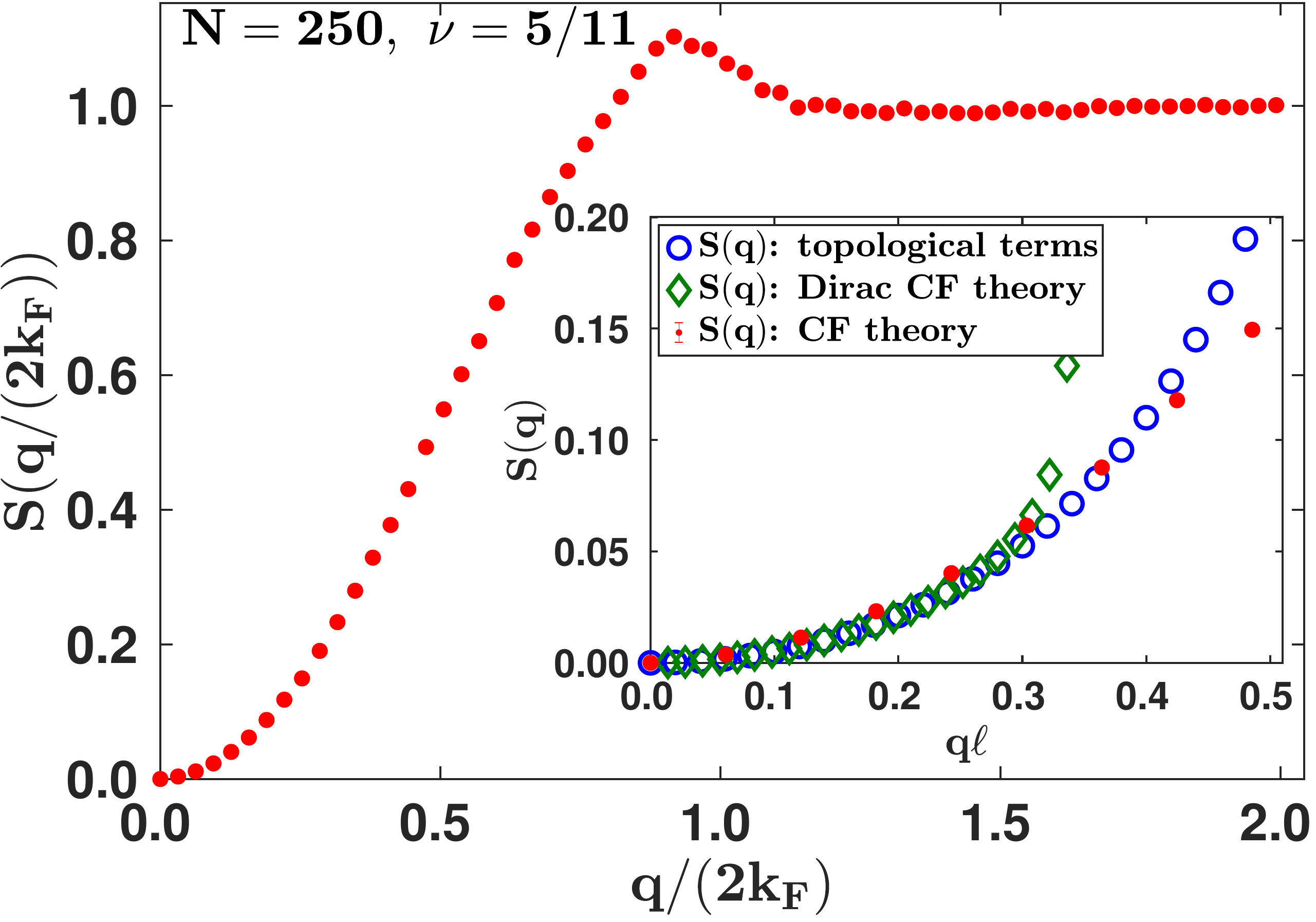}
\includegraphics[width=0.23\textwidth]{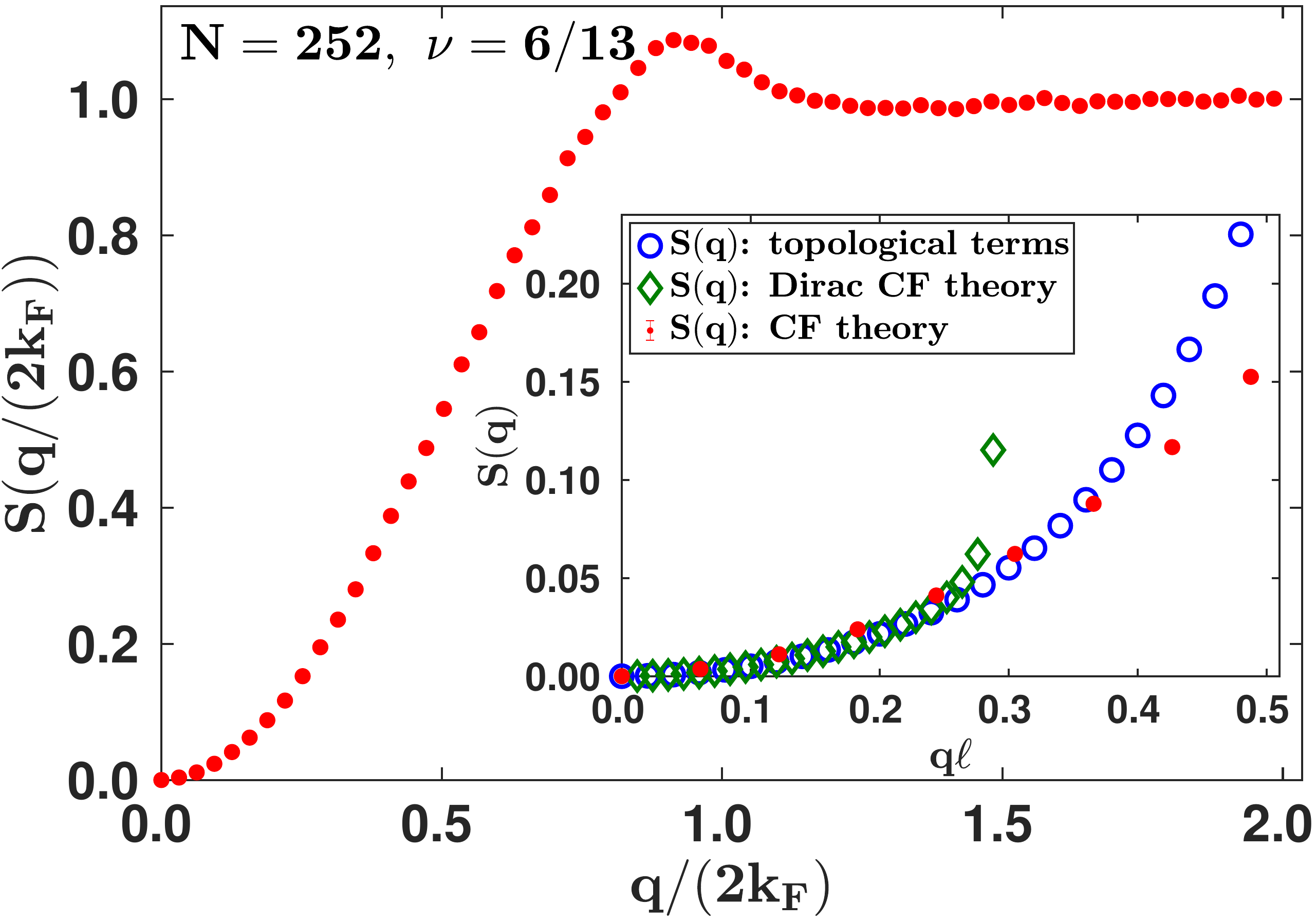} \\ 
\includegraphics[width=0.23\textwidth]{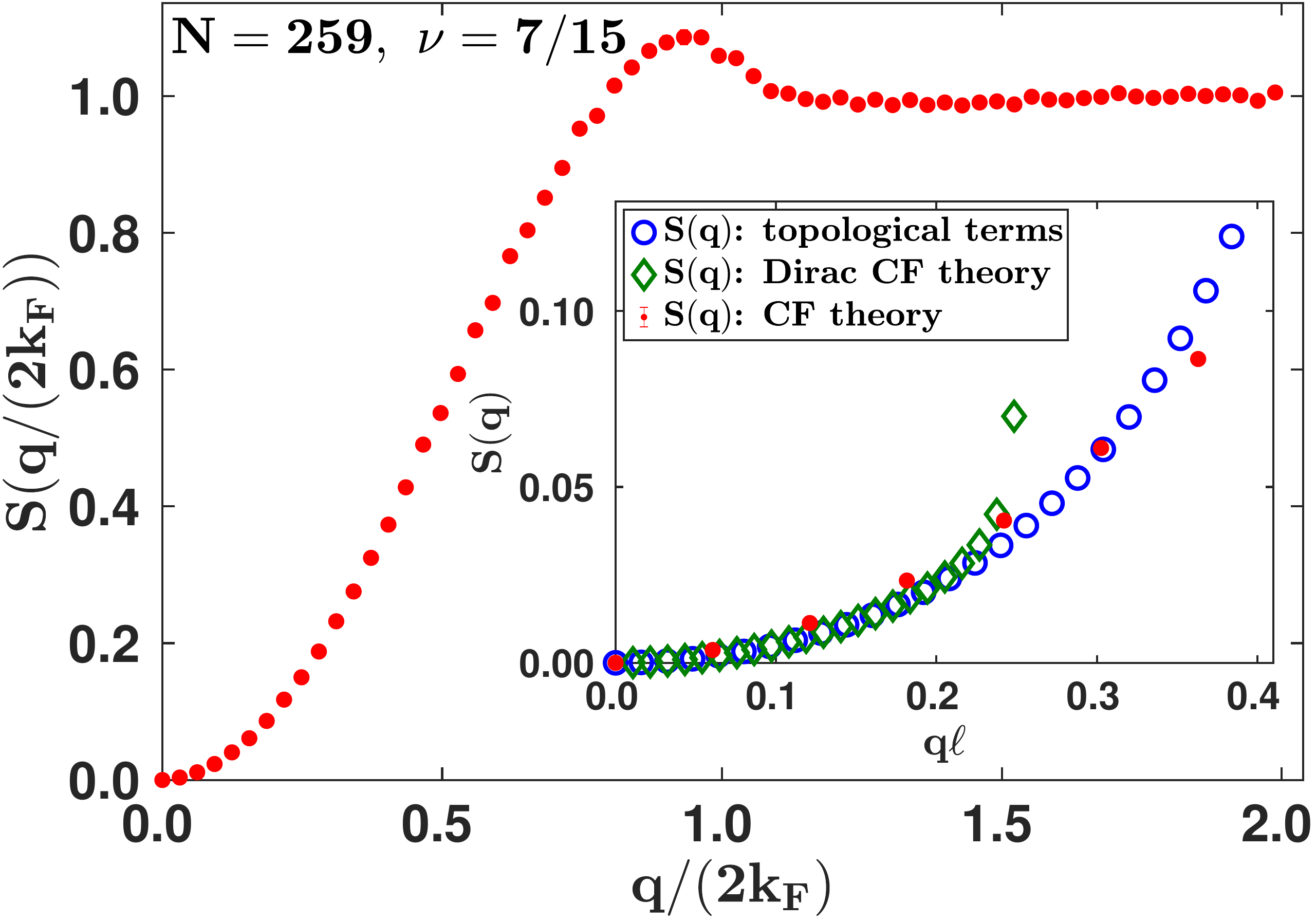}
\includegraphics[width=0.23\textwidth]{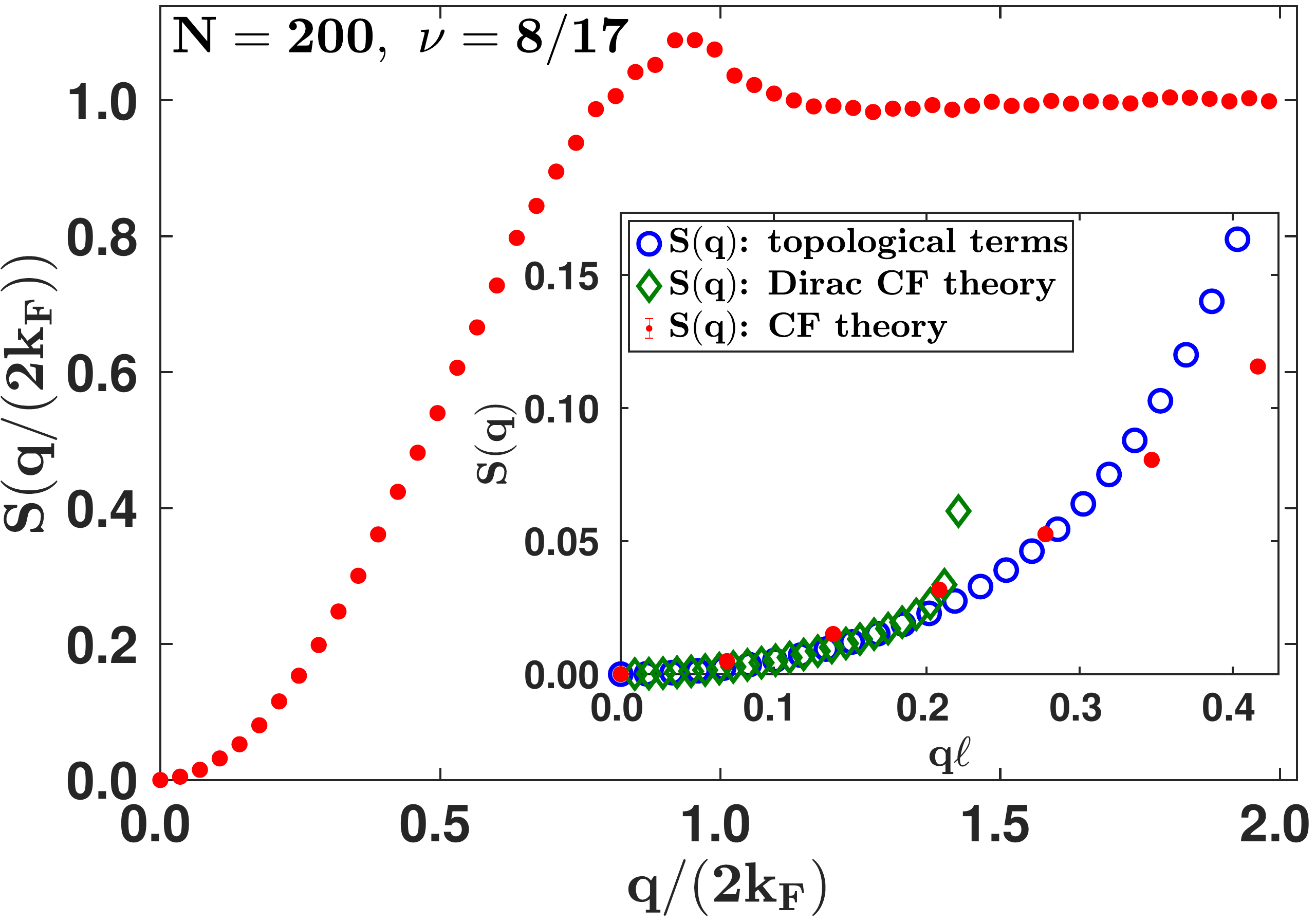} \\ 
\includegraphics[width=0.23\textwidth]{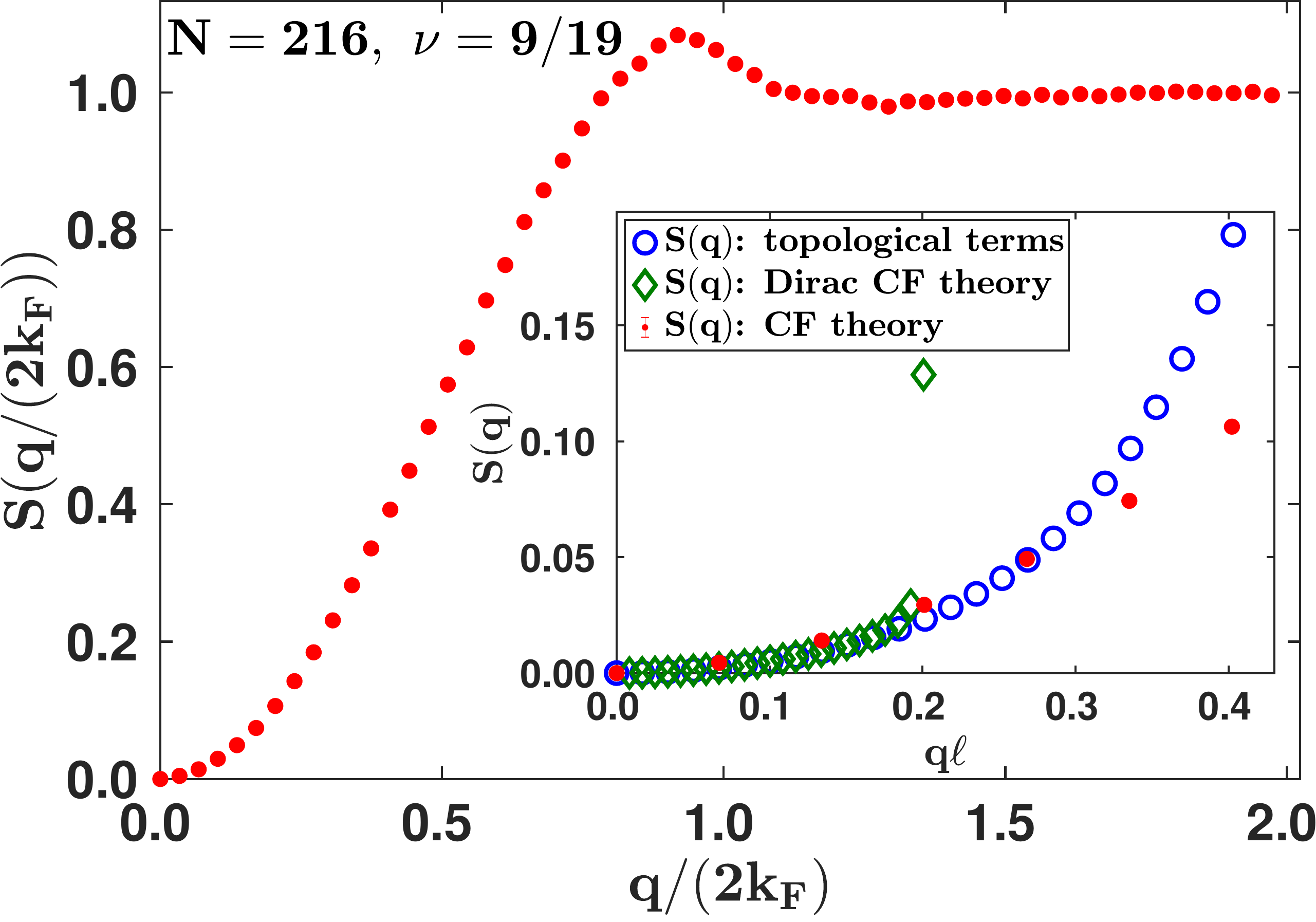} 
\includegraphics[width=0.23\textwidth]{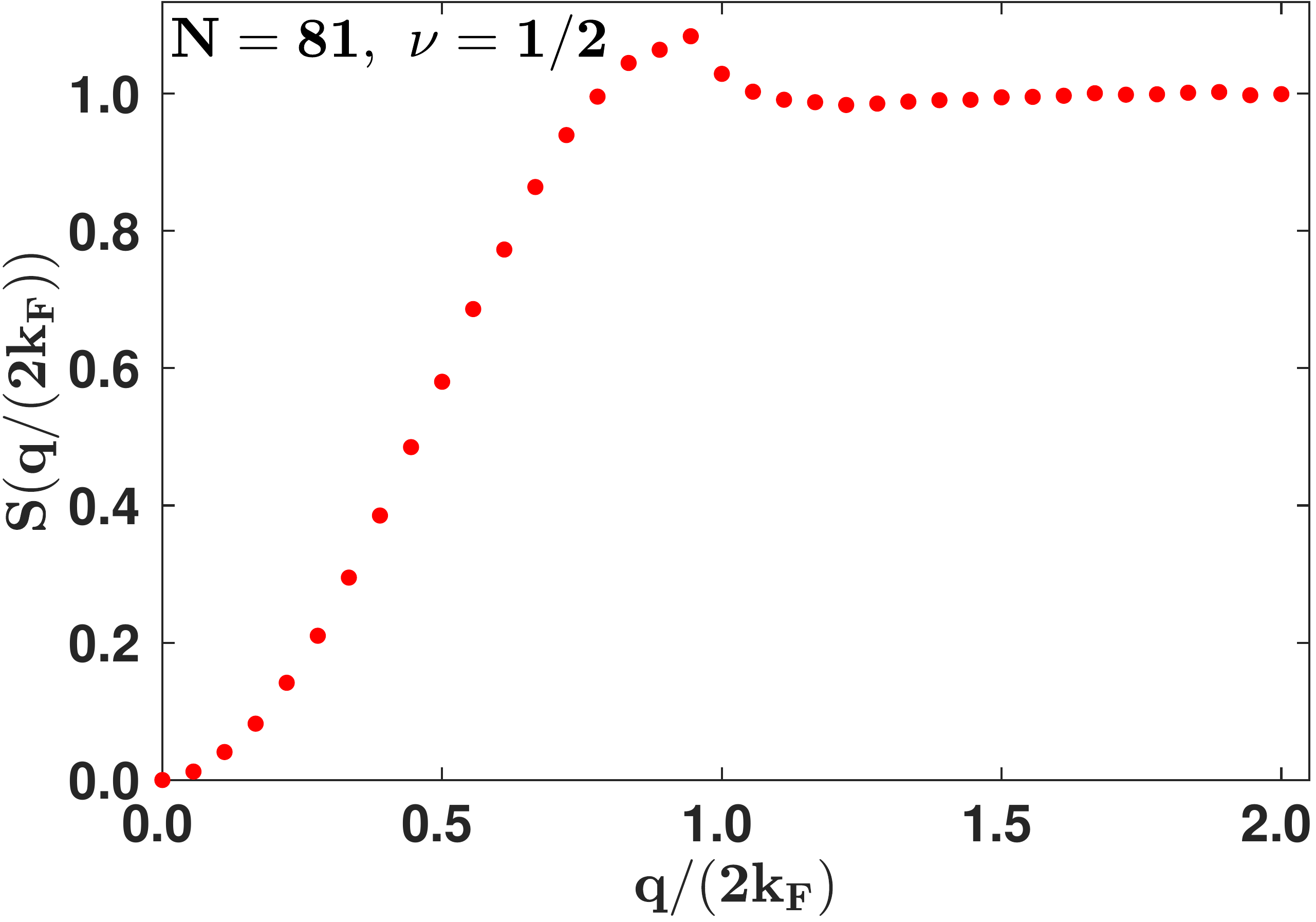} 
\caption{(color online) Static structure factor obtained in the spherical geometry using Eq.~(\ref{eq_Sq_sphere}) with the composite fermion wave function (red dots). The inset shows a comparison in the small wave vector limit with predictions of the topological terms [Eq.~(\ref{eq_structure_factor_top_terms})] (blue circles) and Dirac composite fermion theory [Eq.~(\ref{eq_structure_factor_Dirac_CF})] (green diamonds).}
\label{fig:structure_factor}
\end{center}
\end{figure}

We can also evaluate the structure factor directly from its definition given in Eq.~(\ref{def_Sq}). The magnitude of the planar wave vector $q$ is related to the total orbital angular momentum $L$ on the sphere by the relation: $L=qR$ where $R=\sqrt{Q}\ell$ is the radius of the sphere and the static structure factor $S_{L}\equiv S_{qR}$ is given by\cite{Kamilla97}:
\begin{equation}
 S_{L}=\begin{cases}
       0,~L=0 \\
       \frac{4\pi}{N}\langle |\sum_{j} Y_{L,0}(\boldmath{\Omega_{j}})|^2 \rangle=
       \sum\limits_{i,j} \frac{\left\langle P_{k}\left( \cos \left( \frac{r_{ij}}{R} \right) \right)\right\rangle}{N},~L>0 \\
      \end{cases}
\label{eq_Sq_sphere}      
\end{equation}
where $Y_{l,m}(\boldmath{\Omega}\equiv(\theta,\phi))$ are spherical monopole harmonics with $\theta$ and $\phi$ the polar and azimuthal angles on the sphere, $P_k(x)$ is the $k^{\rm th}$ ordered Legendre polynomial, and $r_{ij}$ is the arc distance between electrons $i$ and $j$ on the sphere. In the above equation we have chosen $L_{z}=0$ without loss of generality since we are only interested in uniform homogeneous states. In Fig.~\ref{fig:structure_factor} we show the static structure factor calculated using Eq.~(\ref{eq_Sq_sphere}) for a large system along the sequence $n/(2n+1)$ and at $\nu=1/2$. \\

Gromov \emph{et al.} found, under certain assumptions, that the static structure factor in the $q\ll 1$ ($\ell=1$) limit for the Jain states is given by\cite{Gromov15,Nguyen17}:
\begin{equation}
S^{\rm top}_{\frac{n}{2n+1}}(q)=\frac{1}{2}q^{2}+\frac{n}{8}q^{4}+\left(\frac{n^{3}+2n^{2}-2n-1}{48} \right)q^{6}+\cdots
\label{eq_structure_factor_top_terms}
\end{equation}
where in the terms corresponding to $q^4$ and $q^6$ can be related to various topological properties of the system. \\

Using the Dirac composite fermion theory\cite{Son15,Nguyen17,Nguyen17b} $S_{n/(2n+1)}(q)$ can be derived exactly in the large $n$ limit, where $q(2n+1)\sim 1$. The static structure factor $S(q)$ in this limit is given by:
\begin{widetext}
\begin{equation}
S^{\rm Dirac}_{\frac{n}{2n+1}}(q)=\frac{[q(2n+1)]^3[(4n+2)^{2}-[q(2n+1)]^2]J_{2}([q(2n+1)])}{32n(2n+1)^{4}J_{1}([q(2n+1)])}+1-e^{-\frac{q^{2}}{2}} 
\label{eq_structure_factor_Dirac_CF}
\end{equation} 
\end{widetext}
where $J_{\alpha}(z)$ is the Bessel function of the first kind. In the $n\rightarrow \infty$ limit (and consequently small $q$ limit) $S^{\rm Dirac}_{n/(2n+1)}(q)$ is identical to $S^{\rm top}_{n/(2n+1)}(q)$. We find that the calculated structure factor agrees well with both $S^{\rm top}_{n/(2n+1)}(q)$ and $S^{\rm Dirac}_{n/(2n+1)}(q)$ in the regime where $q\ell \lesssim 0.1$ (see Fig.~\ref{fig:structure_factor}).

\end{appendices}

\bibliography{biblio_fqhe}
\bibliographystyle{apsrev}
\end{document}